\documentclass[letterpaper,aps,prd,preprint,showpacs,nofootinbib,superscriptaddress]{revtex4-1}
\usepackage{epsf}
\usepackage{epsfig}
\usepackage{graphics}
\usepackage{feynarts}
\usepackage{amssymb}
\usepackage{color}

\newcommand{\nc}{\newcommand}
\nc{\postscript}[2]{\setlength{\epsfxsize}{#2\hsize}\centerline{\epsfbox{#1}}}

\def\ra{\rightarrow}

\newcommand{\Cha}{ {\widetilde{\chi}} }
\newcommand{\Neu}{ {\widetilde{\chi}} }
\newcommand{\al}{{\alpha}}
\newcommand{\wt}{\widetilde}
\newcommand{\hats}{ {\hat s}}
\newcommand{\hatt}{ {\hat t}}

\nc{\beq}{\begin{equation}}   \nc{\eeq}{\end{equation}}
\nc{\bea}{\begin{eqnarray}}   \nc{\eea}{\end{eqnarray}}
\nc{\baa}{\begin{array}}      \nc{\eaa}{\end{array}}
\nc{\bit}{\begin{itemize}}    \nc{\eit}{\end{itemize}}
\nc{\ben}{\begin{enumerate}}  \nc{\een}{\end{enumerate}}
\nc{\bce}{\begin{center}}     \nc{\ece}{\end{center}}
\def\ga{\mathrel{\raise.3ex\hbox{$>$\kern-.75em\lower1ex\hbox{$\sim$}}}}
\def\la{\mathrel{\raise.3ex\hbox{$<$\kern-.75em\lower1ex\hbox{$\sim$}}}}
\nc{\non}{\nonumber}

\begin{document}
\title{
\bf Pair production of neutralinos and charginos at the LHC: 
the role of Higgs bosons exchange 
}
\author{ Abdesslam Arhrib$^{1,2,3}$, Rachid Benbrik$^{3,4,5}$, Mohamed
  Chabab$^{3}$ and Chuan-Hung Chen$^{6,}$ }

\affiliation{
Department of Physics, National Cheng-Kung
University, Tainan 701, Taiwan\\
$^{2}$Facult\'e des Sciences et Techniques, B.P 416 Tangier, Morocco.\\
$^{3}$LPHEA, FSSM, Cadi Ayyad University, B.P. 2390, Marrakesh, Morocco.\\
$^{4}$Instituto de F\'isica de Cantabria (CSIC-UC), Santander, Spain.\\
$^{5}$Facult\'e Polydisciplinaire de Safi, Sidi Bouzid B.P 4162, 46000 Safi,
Morocco.\\
$^{6}$National Center for Theoretical Sciences, Hsinchu 300, Taiwan.
}


\date{\today}
\begin{abstract}
We analyze the effects of the s-channel 
Higgs bosons exchange on the charginos and neutralinos-pair
production in proton-proton collision at the CERN Large Hadron
Collider (LHC) in the following channels: 
$pp \to \Cha^+\Cha^-/\Neu^0\Neu^0 + X$, 
within the minimal supersymmetric standard model (MSSM). 
Assuming the usual GUT relation between $M_1$ and $M_2$ at the weak scale, 
we found that substantial enhancement can be obtained
through $s$-channel Higgs bosons exchange 
in the mixed regime where $M_2 \sim |\mu|$ with
moderate to large $\tan\beta$ at the resonance of the heavy Higgs bosons.
By Combining the phenomenological constraints on neutralinos
and charginos, we may still find regions of parameter 
space where charginos and neutralinos-pair
production at the LHC from $b\bar b$ initial state can be large and
observable at LHC.
We also compute the full complete set of electroweak (EW) 
contributions to $pp\to gg\to \Cha^+\Cha^-/\Neu^0\Neu^0 + X$ at one loop level
 in the general MSSM.
The analytical computation of the complete tree level amplitude for 
$b\bar{b} \to \Cha^+\Cha^-/\Neu^0\Neu^0 + X$, 
including s-channel Higgs exchange,  is given. 
\end{abstract}

\maketitle

\section{Introduction}

The Standard Model (SM)~\cite{Glashow:1961tr,Higgs:1964ia,Englert:1964et}, 
a theory of strong and electroweak interactions, is amazingly 
consistent with most precision measurements up to the present accessible
energies. Nevertheless, the notorious hierarchy problem indicates
that  the SM should be an effective theory at electroweak scale. One
of the solutions to the hierarchy problem is to introduce supersymmetry
(SUSY), where the quadratic divergences induced by one-loop corrections to
Higgs mass are smeared. Therefore, the important extension of the SM in the
framework of SUSY is the minimal supersymmetric standard model (MSSM). If we further impose a discrete 
R-parity $R_p=(-1)^{2S+3(B+L)}$
\cite{Barger:1993cf,Ellis:1983ew,Martin:1992mq,Diehl:1994ff,Kuchimanchi:1993jg}
to the system, 
where the super particles carry odd R-parity and $S$, $B$ and $L$ 
denotes the spin, baryon and lepton number of a particle, respectively, 
a stable lightest supersymmetric particle (LSP) exists and the 
super-partners of the SM particles are always produced in pairs. 

Motivated by the existence of dark matter (DM) that has the abundance of 24\%
in the universe, the neutral stable 
LSP might be considered as DM candidate ~\cite{Davis:1968cp}. 
Although sneutrino, the super-partner of
neutrino, could be a viable candidate of DM, enormous studies are
concentrated on neutralino, where the state consists of neutral gauginos and
higgsinos~\cite{Haber:1984rc}. The interest to adopt neutralino as LSP in the
MSSM is that the corresponding mass matrix in interaction eigenstates only
depends on four unknown parameters and they are $M_{1, 2}$, $\mu$ and
$\tan\beta=v_2/v_1$, where $M_{1[2]}$ is soft SUSY breaking gaugino mass of
$SU(1)[(2)]$ gauge symmetry, $\mu$ is the mixing coefficient of doublets
$\phi_u$ and $\phi_{d}$ in Higgs potential and $v_{1(2)}$ is the vacuum
expectation value (VEV) of $\phi_{d(u)}$. Hence,  if the neutralino is
observed, it not only confirms SUSY, but also provides the clue of DM. 
Additionally, due to the similarity in involved parameters, the possible next
LSP could be chargino, which consists of charged gauginos and higgsinos. 
For completeness, in this paper we study various mechanisms for the 
production of charginos and neutralinos at the Large Hadron 
Collider (LHC) in detail. 

In the literature, the studies of chargino/neutralino pair 
production in the MSSM are concentrated on the Drell-Yan process of
quark-antiquark annihilation and  gluon-gluon fusion. For instance,
the direct production of charginos and neutralinos  at Tevatron/LHC by
$p\bar{p}/pp\rightarrow \tilde{\chi}_{i}\tilde{\chi}_{j} +X$ through
quark-antiquark annihilation at the next-to-leading order (NLO) 
was investigated by Beenakker {\cal et al}~\cite{Beenakker:1999xh}.
The charginos and neutralinos pair production by 
 gluon-gluon fusion were analyzed in Ref.~\cite{Ma:1999ima,Yi:2000dt} in the
 framework of mSUGRA model.  The
neutralino pair production via quark-antiquark annihilation at LHC
was considered by Han {\cal et al.}~\cite{Han:2000ck}. Moreover,
the correlation of beam polarization and gaugino/higgsino mixing was
studied in Ref.~\cite{Debove:2008nr}. It is worth 
mentioning that although chargino/neutralino pair production by
gluon fusion is loop effects,  due to the high luminosity of LHC, 
the production rate can be still significant. 
One can also access to chargino and neutralino pairs from 
Heavy Higgs bosons which could be copiously produced at LHC  
 and followed by their subsequent decays into chargino 
and neutralino pairs. Detail studies of such scenario have been 
adressed in \cite{Bisset:2007mi, Baer:1992kd, Moortgat:2001pp}.

Beside the channels mentioned earlier, in this paper we are
going to explore the case when  the value of $\tan\beta$ is as large as that
of  $m_t/m_b$ and the production mechanism is through the annihilation of  
bottom-antibottom pair  with scalar Higgs ($H^0$, $A^0$) as the
mediator\footnote{Similar
analysis has been done for squark pair production at LHC 
\cite{Arhrib:2009sb} and stau production at hadron colliders \cite{Lindert:2011td}}. 
The reason to study such effect is because the involved coupling is associated with $m_b \tan\beta /v$ and  $v=\sqrt{v^2_1+v^2_2}$. Although the parton distribution function (PDF)
of bottom quark inside proton is smaller than that of light quark, interestingly  the chargino/neutralino production rate will be enhanced naturally in the scenario of large $\tan\beta$.
Furthermore, we also find that another enhanced effect will be created  when the mediated Higgs is tuned to be a resonant Higgs, i.e. the condition $\sqrt{p^2_b + p^2_{\bar b}}=\sqrt{\hat s}\approx m_{H^0,A^0} \approx 2\,m_{\Cha}$
is satisfied. 
Intriguingly,  the same resonant effect plays a 
prominent role in the neutralino DM, where the LSP neutralino 
yields the desired amount of relic density  in some region of the SUSY
parameter space~\cite{Bertone:2004pz}.  

The paper is organized as follows.  
In Sec. II, we introduce the basic properties of charginos and 
neutralinos and the radiative
corrections to the bottom Yukawa coupling  in the MSSM.
In Sec. III, we present the production mechanisms for 
chargino/neutralino pair production via 
 quark  annihilation and gluon fusion
and discuss the 
constraints on the SUSY parameters. We do the detailed numerical analysis on
the production cross sections in Sec. 
IV. We give conclusions in Sec. V. Additionally, 
the relevant couplings of  the chargino/neutralino to gauge 
bosons and Higgs bosons are given in Appendix A. The 
analytic expressions for chargino/neutralino pair production in the exchange of
Higgs  boson are summarized in Appendix B.

\section{ Masses and Yukawa couplings of  charginos and neutralinos}
For studying the production of charginos and neutralinos,  we introduce
the relevant properties of charginos and neutralinos in 
this section, whereas the details of the couplings of 
charginos/neutralinos to gauge bosons, Higgs bosons, 
 fermions and sfermions are given in Appendix A.
For comparing with the results in the literature, 
hereafter,  we adopt the notation that was used in
Refs.~\cite{Haber:1984rc, Gunion:1984yn}. 

\subsection{ Masses of charginos and neutralinos} 

In terms of two-component Weyl spinors, the chargino mass term in the Lagrangian could be described by
\begin{eqnarray}
{\cal{L}}^m_{\Cha^\pm} = -\frac{1}{2} \left( \psi^+ \psi^-\right)
\left( \begin{array}{cc} 0 & {\cal M}^T_C \\ {\cal M}_C & 0 \end{array} \right) \left( \begin{array}{c} \psi^+ \\ \psi^- \end{array} \right) + {\rm H.c}\,, 
\end{eqnarray}
where ${\cal M}_C$ is given by~\cite{Gunion:1984yn} 
 \begin{eqnarray}
{\cal M}_C = \left( \begin{array}{cc} M_2 & \sqrt{2}M_W s_\beta
\\ \sqrt{2}M_W c_\beta & \mu \end{array} \right)
 \end{eqnarray}
with $s_\beta (c_\beta) \equiv \sin\beta ( \cos\beta)$ and  the representations of $\psi^{\pm}_j$ for winos and charged higgsinos are 
 \begin{eqnarray} 
 \psi^{+}_j = \left( -i\lambda^+,  \psi^1_{H_2}\right), \quad \psi^{-}_j = \left( -i\lambda^-,  \psi^2_{H_1}\right), \quad j = 1,2.
 \end{eqnarray}
Since the matrix ${\cal M}_C$  is  not symmetric, 
for diagonalizing it, we need to introduce two   $2\times 2$ unitary  matrices $U$ and
$V$, i.e.
 \begin{eqnarray}
U^* {\cal M}_C V^{-1}  = {\rm diag(m_{\Cha^\pm_1}, m_{\Cha^\pm_2})} \ \ 
\ra \ \ U={\cal O}_- \ {\rm and} \ \ V = 
\left\{
\begin{array}{cc} {\cal O}_+ \ \ \ & {\rm if \ det}{\cal M}_C >0 \,,  \\
            \sigma_3  {\cal O}_+ \ \ \ & {\rm if \ det}{\cal M}_C <0 \,. 
\end{array}
\right. 
 \end{eqnarray}
Here, the third Pauli matrix $\sigma_3$ is used to 
make the eigenvalues of ${\cal M_C}$ to be 
positive and ${\cal O}_\pm$ are the $2\times 2$ rotational 
matrices in which the mixing angles are
\begin{eqnarray}
\tan 2 \theta_- =  \frac{ 2\sqrt{2}M_W(M_2 c_\beta
+\mu s_\beta)}{ M_2^2-\mu^2-2M_W^2 c_{2\beta}} \, , \ \ 
\tan 2 \theta_+ = \frac{ 2\sqrt{2}M_W(M_2 s_\beta
+\mu c_\beta)}{M_2^2-\mu^2 +2M_W^2 c_{2\beta}} \,.
\end{eqnarray}
Accordingly, the mass eigenstates of charginos could be expressed by
\begin{eqnarray}
\Cha^+_{i} = V_{ij}\psi^+_{j}, \quad \Cha^-_{i} = U_{ij}\psi^-_{j}
\end{eqnarray}
and the  corresponding mass eigenvalues are given by
\begin{eqnarray}
m^2_{\Cha_{1,2}^\pm} = \frac{1}{2} \Bigg[ M_2^2+\mu^2+2M_W^2
\mp \sqrt{(M_2^2-\mu^2)^2+4 M_W^2( M_W^2 c^2_{2\beta} + M^2_2+\mu^2
+2M_2\mu s_{2\beta})} \Bigg]\,.
\nonumber 
\end{eqnarray}
If the lightest chargino mass $m_{\Cha^\pm_1}$ is known, $|\mu|$ can be regarded 
as a function of $M_2$ and the angle $\beta$. In the limit $|\mu| \gg M_2,\, M_W$, the masses of charginos could be simplified as
\begin{eqnarray}
m_{\Cha_{1}^\pm}  \simeq   M_2 - \frac{M_W^2}{\mu^2} 
\left( M_2 +\mu s_{2\beta} \right) \,, \ \ 
m_{\Cha_{2}^\pm}  \simeq  |\mu| + 
\frac{M_W^2}{\mu^2} {\rm sign(\mu)} \left( M_2 s_{2 \beta} +\mu \right) \,.
\end{eqnarray}
Clearly, if $|\mu| \ra \infty$, the light chargino corresponds to a 
pure wino state with  $m_{\Cha_{1}^\pm} \simeq M_2$, while the 
heavy chargino corresponds to a pure higgsino state with  
$m_{\Cha_{2}^\pm} = |\mu|$.

Next, we turn to discuss 
the case of the neutralinos.  
Since there are four neutral Weyl spinors, the mass term of 
neutralinos in the Lagrangian is written as
\begin{eqnarray}
{\cal{L}}^m_{\Cha^0} = -\frac{1}{2} \big(\psi^0_i\big)^T 
[{\cal M}_N]_{ij}\psi^0_j + {\rm h.c.}
\end{eqnarray}
with
\begin{eqnarray}
\psi^0_i = (-i\lambda_{\gamma}, -i\lambda_{Z}, 
\psi^1_{H_1} \cos\beta - \psi^2_{H_2} \sin\beta,\psi^1_{H_1} 
\sin\beta + \psi^2_{H_2} \cos\beta), \quad i = 1,...,4,
\end{eqnarray}
where the Weyl spinor in above equation in turn is the photino, 
the zino and the neutral higgsinos. The matrix form of ${\cal M}_{N}$ is 
explicitly given  by

\begin{eqnarray}
{\cal M}_N = \left( \begin{array}{cccc}
M_1 & 0 & -M_Z s_W c_\beta & M_Z  s_W s_\beta \\
0   & M_2 & M_Z c_W c_\beta & -M_Z  c_W s_\beta \\
-M_Z s_W c_\beta & M_Z  c_W c_\beta & 0 & -\mu \\
M_Z s_W s_\beta & -M_Z  c_W s_\beta & -\mu & 0
\end{array} \right)
\end{eqnarray}
with $s_W (c_W) \equiv \sin\theta_W (\cos\theta_W)$ and $\theta_W$ 
being Weinberg angle. 
Since neutralinos are Majorana type fermions, the mass matrix ${\cal M}_{N}$
can be diagonalized by using only one unitary matrix $Z$. 
If we set the physical mass of neutralino $m_{\Cha^0_i}$, 
then the $4\times 4$ unitary matrix Z should satisfy~\cite{Gunion:1984yn}
 \begin{eqnarray}
Z^* {\cal{M}}_{N} Z^{-1} = {\rm diag}(m_{\Cha^0_1}, 
m_{\Cha^0_2},m_{\Cha^0_3},m_{\Cha^0_4})\,.
 \end{eqnarray}
Consequently, the relation between weak and physical eigenstates can be expressed as   
\begin{eqnarray}
\Cha^0_{n} = Z_{ni}\psi^0_i \,.
\end{eqnarray}

Because the complete relation between $m_{\Cha^0_i}$ and the parameters
$M_{1,2}$, $\mu$ and $s_{W}(c_W)$ is complicated, the detailed 
expressions can be found in Ref.~\cite{ElKheishen:1992yv}. Nevertheless, if we take $|\mu|\gg M_{1,2}, M_Z  $, the 
relations can be simplified as~\cite{R10} 
\begin{eqnarray}
m_{\Cha_{1}^0} &\simeq& M_1 - \frac{M_Z^2}{\mu^2} \left( M_1 +\mu s_{2\beta}
\right) s_W^2 \,, \non \\
m_{\Cha_{2}^0} &\simeq& M_2 - \frac{M_Z^2}{\mu^2} \left( M_2 +\mu s_{2 \beta}
\right) c_W^2\,,  \non \\
m_{\Cha_{3}^0} &\simeq& |\mu| + \frac{1}{2}\frac{M_Z^2}{\mu^2} \epsilon_\mu 
(1- s_{2\beta}) \left( \mu + M_2 s_W^2+M_1 c_W^2 \right)\,,  \non \\
m_{\Cha_{4}^0} &\simeq& |\mu| + \frac{1}{2}\frac{M_Z^2}{\mu^2} \epsilon_\mu 
(1+s_{2\beta}) \left( \mu - M_2 s_W^2 - M_1 c_W^2 \right)\,. \label{eq:m_N}
\end{eqnarray}
We see clearly that the first two light neutralinos $\Cha^0_{1}$ and $\Cha^0_2$ are dominated by gauginos of $SU(1)$ and $SU(2)$, respectively, while the last two heavy neutralinos $\Cha^0_{3,4}$ are aligned  to the states of higgsinos. 

\subsection{Yukawa couplings}

It is now well established  that the coupling of the $b$-$\bar{b}$-$H^0_{k}$
induces a modification of the tree-level relation between the 
bottom quark mass and its Yukawa coupling 
\cite{Hall:1993gn,Carena:1999py,Carena:1994bv,Pierce:1996zz}. 
Those corrections are amplified at large $\tan\beta$.
The  modifications can be absorbed by redefining the bottom Yukawa
coupling as 
\begin{equation}
Y^b = \frac{\sqrt{2} m_b }{v \cos\beta} 
 \rightarrow \frac{\sqrt{2}}{v \cos\beta} \frac{m_b}{1 + \Delta_b}\, 
\approx \frac{\sqrt{2}}{v} \frac{m_b}{1 + \Delta_b} \tan\beta
\end{equation} 
where the second expression is valid for large $\tan\beta$ 
and {the SUSY-QCD corrections lead to}
\begin{equation}
\Delta_b =  \frac{2\alpha_s }{3\pi} \,\mu \,m_{\wt g} 
\, \tan\beta \,I(m_{\wt b_1}, m_{\wt b_2}, m_{\wt g}) 
+ \frac{(Y^t)^2 }{16 \pi^2} \,\mu \,A_t \,\tan \beta \,
  I ( m_{\wt t_1}, m_{\wt t_2}, \mu )
\label{deltab}
\end{equation}
$m_{\wt g}$ denotes the gluino mass, and
the function $I$ is given by
\begin{eqnarray}
I(a,b,c) = \frac{-1}{(a^2-b^2)(b^2-c^2)(c^2-a^2)}
\left(a^2 b^2\ln\frac{a^2}{b^2}+b^2 c^2\ln\frac{b^2}{c^2}+
c^2 a^2\ln\frac{c^2}{a^2} \right) 
\label{ifun}.
\end{eqnarray}
In $\Delta_b$ we only keep the dominant contributions from the gluino-sbottom
and charged-higgsino-stop loops because they are proportional to the strong
coupling and to the top Yukawa coupling, respectively, while neglecting 
those that are proportional to the weak gauge coupling.  Note that $\Delta_b$
is evaluated at the scale of SUSY particles $M_{\rm SUSY}$ where the heavy
particles in the loop decouple, whereas the bottom Yukawa coupling  $Y^b(Q)$ 
is determined by the running $b$-quark mass $m_b(Q)$
at the scale $Q$:
\begin{equation} 
Y^b(Q) = \frac{ \sqrt{2} m_b (Q)}{ v \cos\beta}\, 
\frac{1}{1+\Delta_b \left(M_{\rm SUSY} \right )} \;.
\label{yuk-cor}
\end{equation}
The contributions  to the bottom Yukawa couplings
which are enhanced at large $\tan\beta$ can be included
to all orders by making 
the following 
replacements~\cite{Guasch:2003cv,Carena:2006ai}
\begin{eqnarray}
g_{hbb} & \rightarrow & g_{hbb} \frac{1 - \Delta_b \left(M_{\rm SUSY} \right )/(\tan\beta \tan\alpha)}
{1 + \Delta_b\left(M_{\rm SUSY} \right )} \\
g_{Hbb} & \rightarrow & g_{Hbb} \frac{1 + \Delta_b \left(M_{\rm SUSY} \right )\tan\alpha /\tan\beta}
{1 + \Delta_b\left(M_{\rm SUSY} \right )} \\
g_{Abb} & \rightarrow & g_{Abb} \frac{1 - \Delta_b \left(M_{\rm SUSY} \right )/\tan^2\beta}
{1 + \Delta_b\left(M_{\rm SUSY} \right )} 
\label{yuk-cor1}
\end{eqnarray}
where
\begin{eqnarray}
g_{hbb}&=&\frac{g m_b }{2 m_W }
\frac{\sin\alpha }{\cos\beta }
=-\frac{g m_b}{2 m_W} (\sin(\beta-\alpha) -\tan\beta \cos(\beta-\alpha))\\
g_{Hbb}&=&\frac{g m_b }{2 m_W}
\frac{\cos\alpha }{\cos\beta }=
=
\frac{g m_b}{2 m_W} (\cos(\beta-\alpha) +\tan\beta \sin(\beta-\alpha))\\
g_{Abb}&=&\frac{g m_b }{2 m_W}\tan\beta
\label{tanbenh}
\end{eqnarray}
As we can see from the above equations, all Higgs couplings to the bottom
quarks have some $\tan\beta$ enhancement at large $\tan\beta$ limit. 
Note also that an other $\tan\beta$ dependence comes through 
$\Delta_b$ corrections.\\
We now have all the ingredients to compute 
the chargino and neutralino pair production at the LHC.


\section{Production processes and constraints}
\subsection{$pp\to \tilde\chi_{i} \overline{\tilde\chi}_j$ via 
quark  annihilation and gluon fusion}

\begin{figure}[hpbt]
\includegraphics*[width=6 in]{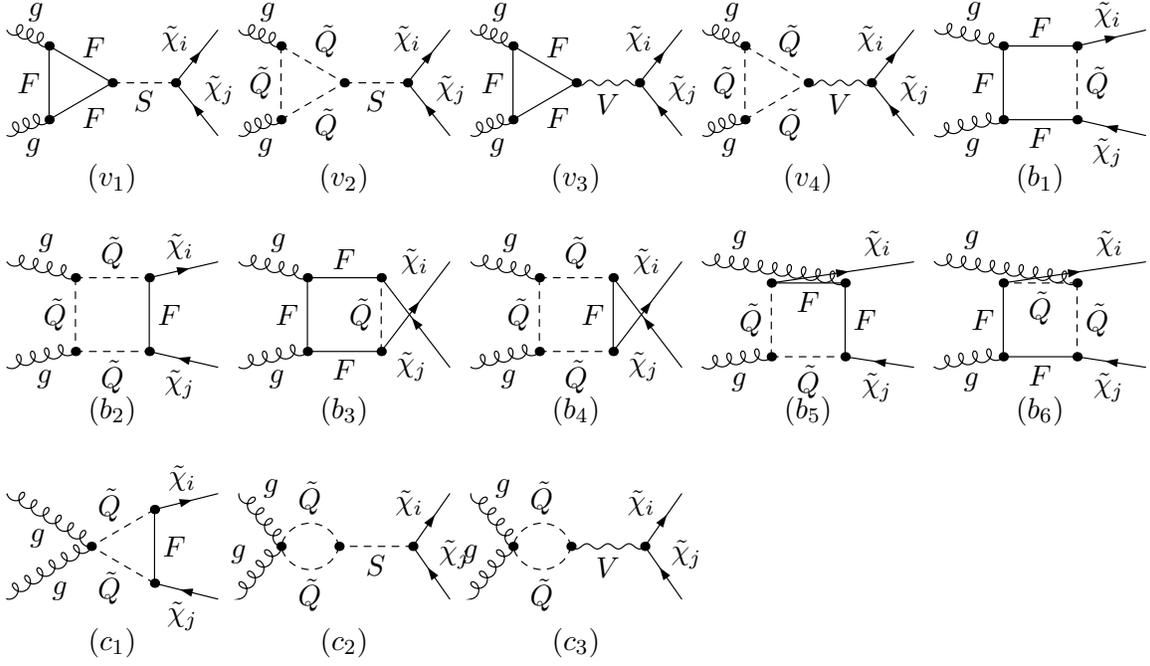}
\caption{One-loop Feynman Diagrams of Chargino-pair production 
at the LHC via gluon-gluon fusion with $S = h^0$, $H^0$ or $A^0$, $V= Z$ 
and $\gamma$ (only if i=j) and $\tilde{Q} = \tilde{u}$ or $\tilde{d}$ 
is squark.}
\label{fig:glglchar-diagrams}
\end{figure}

\begin{figure}[hpbt]
\includegraphics*[width=5 in]{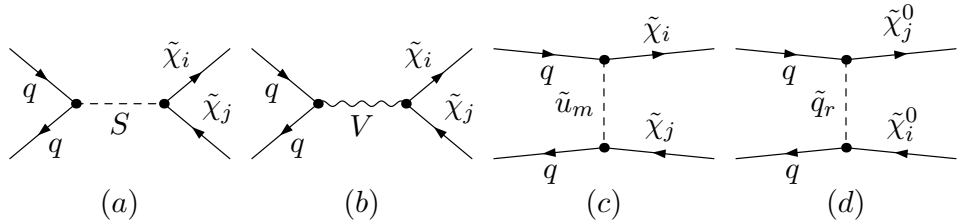}
\caption{ Tree-level Feynman Diagrams of Chargino-pair production 
at the LHC via quark-anti-quark annihilation with $S = h^0$, $H^0$ or
$A^0$, $V= Z$ and $\gamma$ (only if i=j) and $\tilde{u}_m$ is a
squark corresponding to quark q. }
 \label{fig:bbcharg-diagrams}
\end{figure}

As stated early, the colorless fermionic superparticle pair production is
through $gg\to \tilde{\chi}_{i} \overline{\tilde\chi}_j$ and $q\bar q \to
\tilde{\chi}_{i} \overline{\tilde\chi}_j$ channels at hadron colliders. For
gluon-gluon fusion, only loop effects are involved. In terms of type of loop,
we classify the one-loop diagrams into three groups and sketch them in
Fig.~\ref{fig:glglchar-diagrams}; they are:  
(1) triangle diagrams
[Fig.~\ref{fig:glglchar-diagrams}($v_{1}$)-Fig.~\ref{fig:glglchar-diagrams}($v_{4}$)],
(2) box diagrams 
[Fig.~\ref{fig:glglchar-diagrams}($b_{1}$)-Fig.~\ref{fig:glglchar-diagrams}($b_{6}$)]
and (3) the  diagrams with quartic vertices 
[Fig.~\ref{fig:glglchar-diagrams}($c_{1}$)-Fig.~\ref{fig:glglchar-diagrams}($c_{3}$)], where $F$ in the loop denotes the SM quarks, 
$\tilde Q$ is the possible squarks, 
$S$ stands for the scalar bosons ($h^0$, $H^0$, $A^0$) 
in the MSSM and $V$  represents  the gauge bosons  $Z$ and $\gamma$. 
We note that since the electromagnetic interactions are independent of 
the species of $\tilde\chi_i$,  there exist only the 
interactions $\tilde\chi_i$-$\tilde\chi_i$-$\gamma$ (i=1, 2). 
For quark-antiquark annihilation, the leading contributions to $\tilde\chi_i
\overline{\tilde\chi}_j$ production 
are only from the effects of tree level.   The associated
Feynman diagrams are displayed in Fig.~\ref{fig:bbcharg-diagrams}. For
chargino-pair production, the squark $\tilde u_m$ in
Fig.~\ref{fig:bbcharg-diagrams}(c) could be up (down) type squark 
while the squark $\tilde q$
 could be down (up) type squark.  
Although the gluon-gluon fusion loop, s-channel
gauge boson exchange and t-channel squark exchange contributions have been
studied in the literature,  we emphasize that the effects of
Fig.~\ref{fig:bbcharg-diagrams}(a) with $q=b$ and  large $\tan\beta$ on the
$\tilde\chi_i \overline{\tilde\chi}_j$ production 
have not been
explored yet.  Moreover, since the masses of scalar bosons are free parameters, 
when the condition $(p_{\tilde\chi_i}+p_{\bar{\tilde\chi}_j})^2 \approx
m^2_{H^0, A^0}$ is satisfied, the production cross section 
will be enhanced by the resonant Higgs effects. 

By combining the contributions of gluon-gluon fusion and 
quark-antiquark annihilation, 
the cross section for $\Cha_i \overline\Cha_j$ production
in proton-proton collisions at center of mass energy 
$\sqrt{s}$ can be written as
\begin{eqnarray}
\sigma_{\Cha^+_i\Cha^-_j}(s) = \sum_{q} \int^1_{\tau_0} d\tau 
\frac{d{\cal{L}}^{pp}_{q\bar{q}}}{d\tau} \hat{\sigma}_{LO} 
(q\bar{q}\to \Cha^+_i\Cha^-_j)(\tau s) + \int^1_{\tau_0} d\tau 
\frac{d{\cal{L}}^{pp}_{gg}}{d\tau} \hat{\sigma}_{LO} 
(gg\to \Cha^+_i\Cha^-_j)(\tau s)
\end{eqnarray}
with $\tau_0 = (m_{\Cha_i}^2 + m_{\Cha_j}^2 )^2 / s$, and 
the parton luminosity is
\begin{eqnarray}
\frac{d{\cal{L}}^{pp}_{ab}}{d\tau} = \int^1_{\tau} \frac{dx}{x} 
\frac{1}{1+\delta_{ab}} \big[ f_{a}(x,\mu_F) f_{b}(\frac{\tau}{x}, \mu_F) 
+ f_{b}(x, \mu_F) f_{a}(\frac{\tau}{x}, \mu_F)\big ]  
\end{eqnarray}
where $f_{a}(x, \mu_F)$ is parton distribution function (PDF) for parton 
$a$ inside proton and  x 
 is the momentum fraction  at the scale  
$\mu_F = m_{\Cha_i} + m_{\Cha_j}$.

\subsection{Constraints on the free parameters of the MSSM}

For studying the numerical analysis, we need the information of constraints that are from experimental conditions and data and theoretical requirements~\cite{Djouadi:2005an,Dedes:2003cg}. We summarize them as follows:

\begin{itemize}
\item The most stringent constraint generally arises from $\Delta\rho^{SUSY}$
which receives contributions from both stop and sbottom. The extra
contributions to the $\Delta\rho^{SUSY}$ parameter from the stop and
sbottom sector~\cite {Djouadi:1996pa,hagiwara} should not exceed the current
limit from precision measurements \cite{PDG}  i.e.  $\Delta\rho^{SUSY}$
$\leq$ $ 10^{-3}$. Note that this constraint will not affect the parameter space that is associated with the effects of charginos and neutralinos~\cite{hagiwara}. 

\item The soft SUSY-breaking parameters $A_q$ at the weak scale should not
be too large in order to keep the radiative corrections to the Higgs
masses under control.  In particular the trilinear couplings of the
third generation squarks $A_{t,b}$,  will play a particularly
important role in the MSSM squarks/Higgs sectors. These parameters can
be constrained in at least one way, besides the trivial requirement
that it should not make the off-diagonal term of the squark mass
matrices too large to generate too low masses for the
squarks. $A_{t,b}$ should not be too large to avoid the occurrence of
charge and color breaking (CCB) minima in the Higgs 
potential~\cite{Casa:1996ca}.
%

\item Another constraint 
 is  the perturbativity of the
bottom Yukawa coupling $Y^b $. Since the radiative corrections to 
the bottom Yukawa
couplings have been implemented in Eq.~(\ref{yuk-cor}) that 
may blow up when SUSY parameters vary. Thus, we adopt $Y^b \la (4 \pi)^2$.

\item We have imposed also all the experimental bounds on 
squark, chargino, and neutralino masses as well as Higgs boson 
masses \cite{PDG}.

\item We assume that $\Cha^0_1$ is the LSP and will 
escape from the detection.
\end{itemize}


\section{Numerical Analysis and Discussions}
After introducing the physical effects and constraints, 
we now discuss the numerical analysis 
for the inclusive production
cross sections of chargino and neutralino  with $\sqrt{s}=7$ and $14$ TeV 
at the LHC.
Since there are many free parameters in MSSM, for simplifying the study, 
we adopt  the scenario of universal  soft SUSY breaking for the trilinear 
couplings, i.e. $A_t=A_b$, and for the squark masses to be 
$M_{\widetilde{Q}} = M_{\widetilde{U}} \equiv M_{SUSY}$.
Accordingly, the Higgs masses $m_{h^0,H^0,H^\pm}$ and mixing $\alpha$ are 
fixed in terms of the CP-odd mass $m_{A^0}, \tan\beta$ as well as  
$M_{SUSY}$, $A_{b,t}$, $M_2$ and $\mu$
 for higher order corrections \cite{Heinemeyer:1998yj}. 
All the MSSM Higgs masses and 
relevant parameters are computed with FeynHiggs code \cite{Heinemeyer:1998yj}.
We use CTEQ6L parton distribution 
functions~\cite{Nadolsky:2008zw,Brock:1993sz} 
to estimate the various cross sections. 
Moreover, in order to improve the perturbative calculations,
one-loop running mass formula for $m_b(Q)$ is taken by
\begin{eqnarray}
m_b(Q) = m_b^{\rm{\overline{DR}}}(Q) = m_b^{\rm{\overline{MS}}}(Q) 
\left( 1 + \frac{4 \alpha_s}{3\pi}\right)\,,
\end{eqnarray}
where $m_b^{\rm{\overline{MS}}}$ includes the SM QCD corrections and
the running QCD coupling $\alpha_s$ is calculated at the two-loop
level \cite{alfas}.  The
light-quark masses are neglected in the numerical calculations. 
Other values of SM parameters are chosen as
$m_t = 173$ GeV, 
$m_W = 80.398$ GeV,
$m_Z = 91.1878$ GeV and
$m_b(m_b) = 4.25$ GeV \cite{PDG}.
The fine structure constant is taken at the $Z$ pole with
$\alpha_{ew}(m^2_Z) = 1/128$~\cite{PDG}.
For other MSSM parameters, we will perform a systematic scan in the 
following range:
\begin{itemize}
\item 120$\, {\rm GeV} \le m_{A^0} \le 600\, {\rm GeV}$; 

\item 3 $\le \tan\beta \le $40;

\item 100$\,{\rm GeV} \le \mu \le 1\,{\rm TeV}$;\\
The sign of $\mu$ is taken positive, as preferred by the SUSY explanation of
the $(g-2)_{\mu}$ anomaly.

\item 100$\,{\rm GeV} \le M_{2} \le 450\,{\rm TeV}$;\\
We impose the GUT relation at weak scale to fix $M_1$.
\end{itemize}

Before displaying our results, 
we emphasize that 
the MSSM parameter space has been subject to the experimental constraints 
of Tevatron and  LHC by the negative search of
some specific processes. By looking to the
Higgs boson production in tau-tau final states, both Tevatron and CMS 
\cite{Chatrchyan:2011nx,Benjamin:2010xb} have set a limit 
on  ($\tan\beta$, $m_{A^0}$) for some specific scenarios in the 
framework of the MSSM. From CDF and D{\O}  (respectively CMS) data, 
those limits on ($\tan\beta$, $m_{A^0}$) are only valid for $m_{A^0}\la 200$ GeV
(respectively $m_{A^0}\la 300$ GeV). From CMS data $\tan\beta\ge 30$ is already
excluded for $100\la m_{A^0}\la 200$ GeV in the MSSM with 
maximal mixing scenario, while for 
$200\la m_{A^0}\la 300$ GeV the $\tan\beta$ is limited  in the range $[30,\, 55]$.
For our presentation, we will not restrict ourselves with those 
experimental constraints shown in 
Refs.~\cite{Chatrchyan:2011nx,Benjamin:2010xb}
but rather present a complete scan over the MSSM
parameter space. In the mean time, 
in our analysis we restrict ourselves to the $\tan\beta \la 40$ for 
which $m_{A^0}\ge 150$ GeV is allowed. However, according to ATLAS and CMS
analysis \cite{Chatrchyan:2011nx,Benjamin:2010xb} care must be taken 
for low value of $m_{A^0}\approx 150$ GeV where $\tan\beta$ 
should be less than $\approx 25$.


\begin{figure}[t] 
\begin{picture}(320,260)
\put(-100,-250){\mbox{\psfig{file=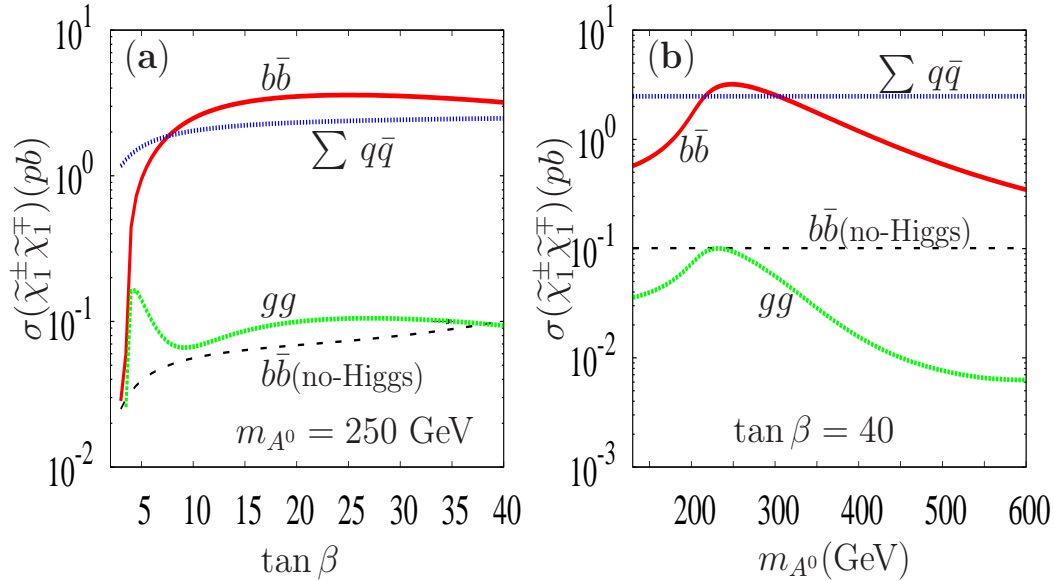,width=8.in,height=11in}}}
\end{picture}
\vspace{-0.1cm}
\caption{Separate cross sections for chargino 
 pair production $\sigma (\Cha^\pm_1 \Cha^\mp_1)(pb)$ 
in pico barn at the LHC with $\sqrt{s} = 14$ TeV as a function 
of $\tan\beta$ (left) and $m_{A^0}$ (right).
The SUSY parameters are chosen to be $M_{SUSY} = 490$ GeV  $M_2, \mu = 120, 150$ GeV,
$A_t = A_b =1140$ GeV.} 
\label{fig-ppchar11}
\end{figure}

\begin{figure}[t] 
\begin{picture}(320,260)
\put(-100,-250){\mbox{\psfig{file=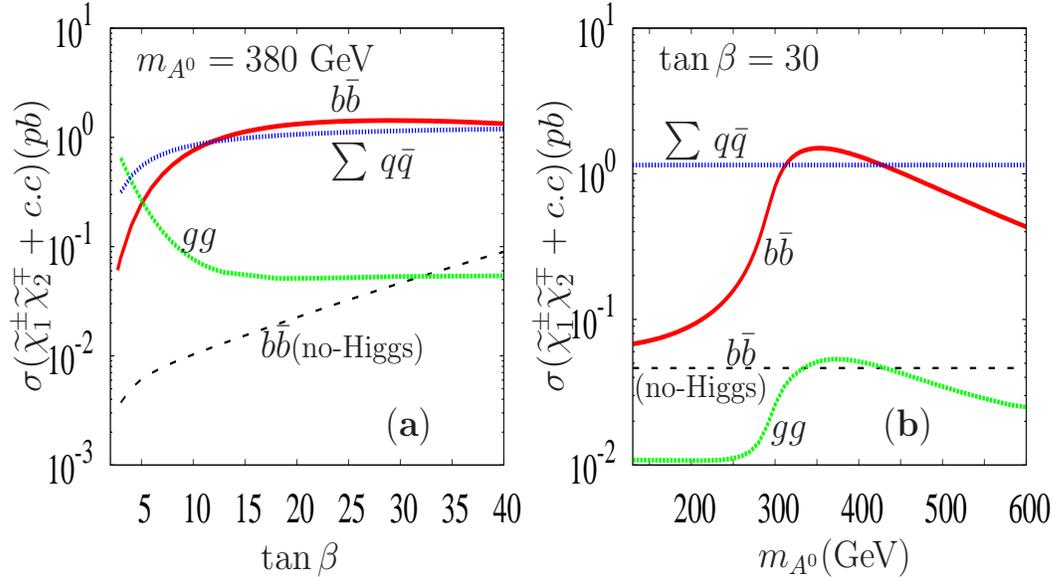,width=8.in,height=11in}}}
\end{picture}
\vspace{-0.1cm}
\caption{Separate cross sections for chargino pair production 
$\sigma(\Cha^+_1\Cha^-_2 + c.c)(pb)$ in pico barn at the LHC with 
$\sqrt{s} = 14$ TeV as a function of $\tan\beta$ (left) 
and $m_{A^0}$ (right).
The SUSY parameters are chosen to be  $M_2, \mu = 120, 150 $ GeV, 
 $A_t = A_b =1140$ GeV.} 
\label{fig-ppchar12-1}
\end{figure}

In order to obtain the correct numerical results,   we first check the  calculations for chargino and neutralino pair
production by gluon-gluon fusion in mSUGRA model. Our results are
qualitatively consistent  with  Ref.~\cite{Ma:1999ima,Yi:2000dt}.\\
For illustration, we show the production cross sections  as a function of $\tan\beta$ [$m_{A^0}$]  at
$\sqrt{s}=14$ TeV  for 
$\sigma(b\bar{b},gg, \sum q\bar{q} \to \Cha^+_1\Cha^-_1)$
 and $\sigma(b\bar{b},gg, \sum q\bar{q} \to 
\Cha^+_1\Cha^-_2+c.c)$ in Fig. \ref{fig-ppchar11}(a)[(b)] and 
Fig.~\ref{fig-ppchar12-1}(a) [(b)], respectively. All the cross sections presented here are only at the 
leading order without K-factor. 
The NLO corrections to chargino/neutralino pair production have
been done in Ref.~\cite{Beenakker:1999xh}, where the K-factor 
is taken by 1.25 (1.40) for $m_{\chi}\approx 250\, (100)$ GeV. 
In order to understand the sensitivities of $m_{A^0}$ and $\tan\beta$, in the figures we show separately the process for producing chargino pair, e.g. 
the curve of $b\bar b$ (no-Higgs) denotes the bottom-induced Drell-Yan
contributions in which the processes include the s-channel photon and $Z$ boson exchange and t-channel with squark exchange. 
As to the curve of $b\bar b$, it stands for all Higgs-mediated effects and has the enhancement of large $\tan\beta$ that we would like to emphasize in this paper. 

Hence, from  Figs.~\ref{fig-ppchar11}(a) and 
\ref{fig-ppchar12-1}(a), it is easy to find that  although at low $\tan\beta$ the production cross section is dominated by the light-quark fusion, however, the contributions from Higgs-mediated effects through $b\bar b$ annihilation will be over the light-quark fusion when $\tan\beta$ is around $10$. The results show not only the sensitivity of production cross section to  $\tan\beta$ but also the importance of $\tan\beta$ in the mechanism of Higgs exchange, i.e.  the Higgs-mediated effects with large $\tan\beta$ could become dominant in chargino pair production.  Beside the $\tan\beta$ enhanced factor, as mentioned earlier,  Higgs-resonance can be another effect to enhance the chargino-pair production.  We can see the enhancement from 
Figs.~\ref{fig-ppchar11}(b) and \ref{fig-ppchar12-1}(b). By the curve arisen from $b\bar b$ fusion,  
it is clear that there is a bump at $m_{A^0}\approx 250[350]$ GeV in Fig.~\ref{fig-ppchar11}(b) [\ref{fig-ppchar12-1}(b)], where the bump is formed when $m_{H^0}\approx m_{A^0} 
\approx 2m_{\Cha^+_1}$ is satisfied. We note that the curve denoted by  $b\bar b$(no-Higgs) is not sensitive to $\tan\beta$ and has no Higgs-resonance, therefore, its contribution is far below that by Higgs-mediated effects.\\
Although gluon-gluon fusion can contribute to  chargino-pair production by loop effects,  its contributions are much smaller than those from $q\bar{q}$ and $b\bar{b}$ fusion, except the case for  $gg\to \Cha^{\pm}_1\Cha^{\mp}_2$ at low $\tan\beta$.  Since we are considering the scenario with large $\tan\beta$, gluon-gluon fusion is not a dominant process. Therefore, we don't further discuss the gluon-gluon fusion in detail. 

\begin{figure}[t] 
\begin{picture}(320,260)
\put(-100,-250){\mbox{\psfig{file=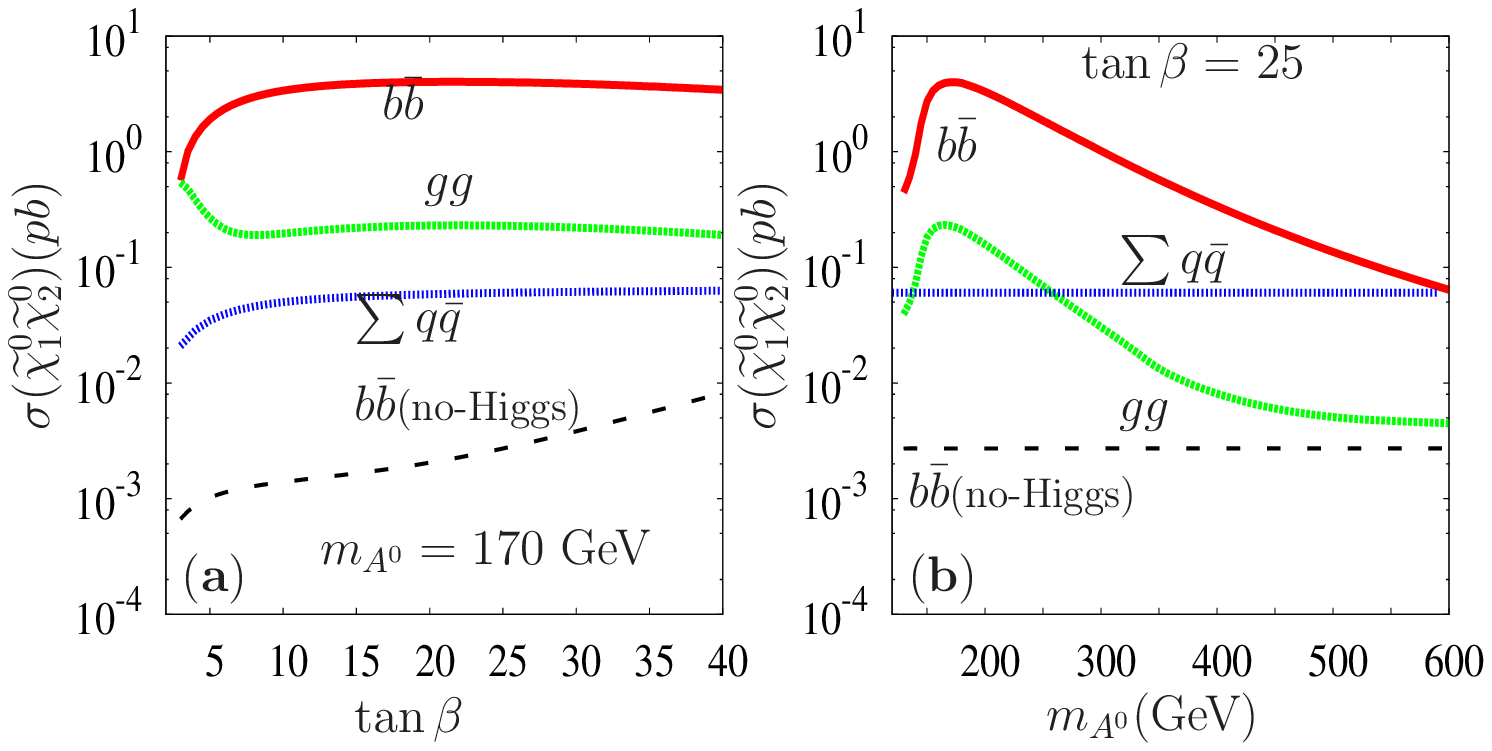,width=8.in,height=11in}}}
\end{picture}
\vspace{-0.1cm}
\caption{Separate cross sections for neutralino pair production  $\sigma(\Cha^0_1\Cha^0_2)(pb)$  at the LHC with $\sqrt{s} = 14$ TeV as a function of 
$\tan\beta$ (left) and as a function of $m_{A^0}$ (right).
The SUSY parameters are chosen to be $M_{SUSY} = 490$ GeV, $M_2 = 120$ GeV, $\mu = 150 $ GeV and $A_t = A_b =1140$ GeV.} 
\label{fig-ppneu12}
\end{figure}

\begin{figure}[t] 
\begin{picture}(320,260)
\put(-100,-250){\mbox{\psfig{file=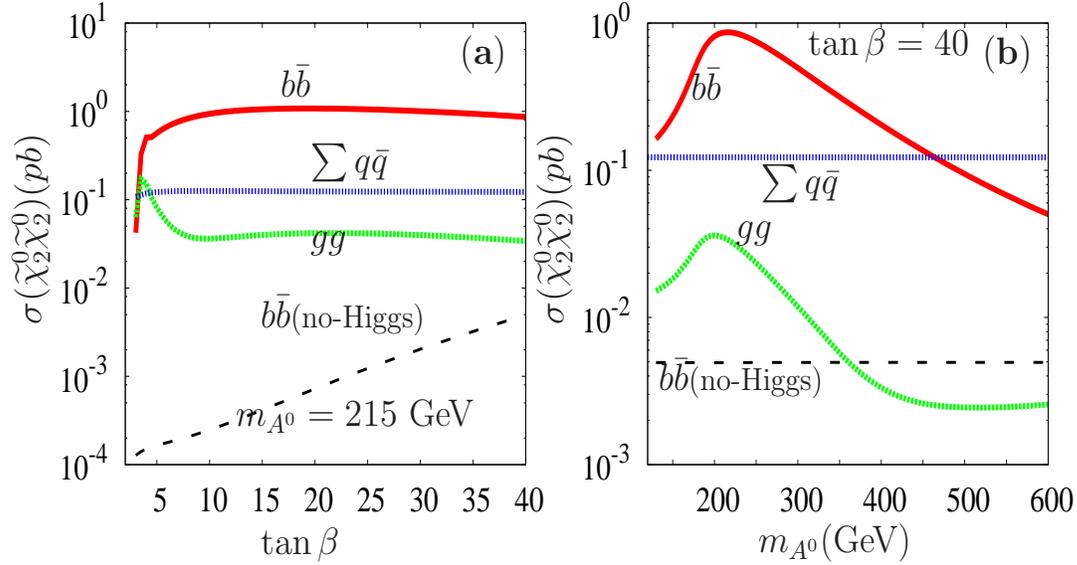,width=8.in,height=11in}}}
\end{picture}
\vspace{-0.1cm}
\caption{Separate cross sections for neutralino pair production  $\sigma(
  \Cha^0_2\Cha^0_2)(pb)$ at the LHC with $\sqrt{s} = 14$ TeV as a 
function of $\tan\beta$ (left) and as a function of $m_{A^0}$ (right).
The SUSY parameters are chosen to be $M_{SUSY} = 490$ GeV, $M_2 = 120$ GeV, 
$m_{\tilde{g}} =  1 $ TeV, $\mu = 150 $ GeV, $A_t = A_b = 1140$ GeV.} 
\label{fig-ppchar113}
\end{figure}

Next, we discuss the situation for neutralino-pair production. Since the lightest neutralino-pair 
is associated with invisible signal, we skip the relevant discussions. Accordingly, we will concentrate on the production of $\chi^{0}_1 \chi^{0}_2$ and $\chi^{0}_2 \chi^{0}_2$ pairs. 
Additionally, the production channels 
$pp\to\Neu^0_1\Neu^0_2$ and $pp\to \Neu^0_2\Neu^0_2$ are of special interest 
 because of the presence of dileptons in their decay products. 
 
Similar to the chargino cases, we show
various  production cross section $\sigma(b\bar{b},gg, \sum q\bar{q} 
\to\Neu^0_1\Neu^0_2 , \Neu^0_2\Neu^0_2) $ 
as a function of $\tan\beta$ [$m_{A^0}$]  in Fig.~\ref{fig-ppneu12} [\ref{fig-ppchar113}] 
for 14 TeV LHC energy. 
In both cases, near the resonance region and for large $\tan\beta$, 
 one can see that $b\bar{b}$ fusion contribution 
is more important  than $q\bar{q}$ contribution and can go up to 
one order of magnitude larger exceeding few pico-barn in some cases. 
This is mainly due to the smallness of $Z\Neu^0_i\Neu^0_j$ coupling which
contributes to $q\bar{q}$ fusion through $Z$ exchange. 
 Moreover, in the mixed $(|\mu| \sim M_2)$
  regime, the first and second generation squarks would be
  significantly heavier than wino like charginos and neutralinos,
  making the t-channel contribution negligible with respect to the
 s-channel contribution which enjoy the resonant effect 
$\sqrt{\hat s}\approx m_{H^0,A^0} \approx 2\,m_{\Cha}$.
It has to be noted also that the gluon gluon fusion $gg\to\Neu^0_i\Neu^0_j$, 
both for diagonal production $\Neu^0_2\Neu^0_2$ as well as for non-diagonal
 one $\Neu^0_1\Neu^0_2$, is in some cases larger that the $q\bar{q}$ 
fusion in the case of low $\tan\beta$.

\begin{figure}[t] 
\begin{picture}(320,260)
\put(-100,-250){\mbox{\psfig{file=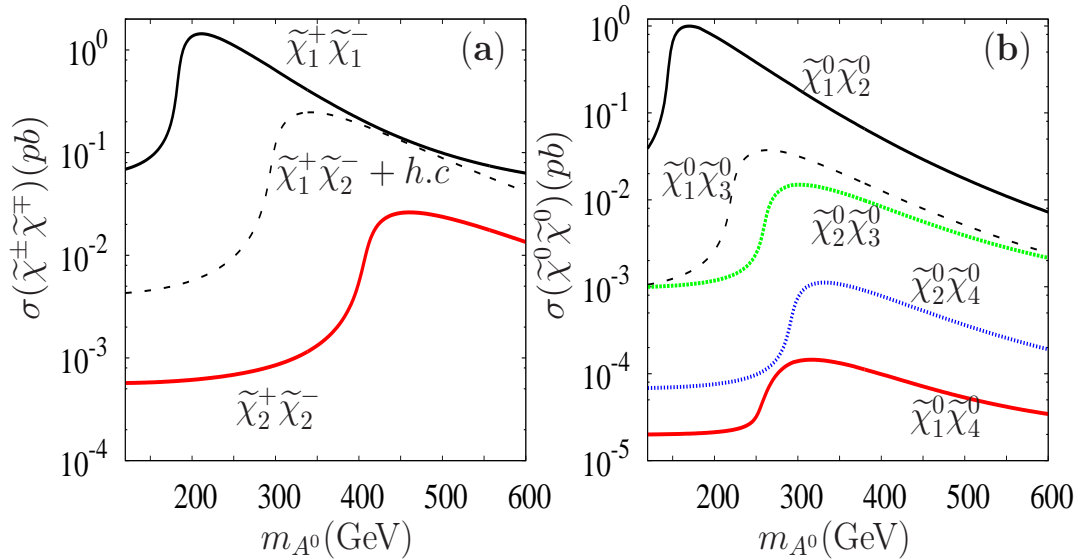,width=8.in,height=11in}}}
\end{picture}
\vspace{-0.1cm}
\caption{Total cross sections for chargino (left) and neutralino (right) pairs production  at the
  LHC with $\sqrt{s} = 7$ TeV as a function of $m_{A^0}$.
The SUSY parameters are chosen to be $M_{SUSY} = 490$ GeV, $M_2 = 120$
GeV, $M_{SUSY} =  490 $ GeV, $\mu = 150 $ GeV, $ A_t = A_b = 1140$ GeV
and is fixed at $\tan\beta = 20$ .} 
\label{fig-ppchar114}
\end{figure}

For comparison, we also present the results for the production 
of chargino and neutralino at $\sqrt{s}=7$ TeV in Fig.~(\ref{fig-ppchar114}).
It is easy to see that large $\tan\beta$ and the Higgs resonant 
effects could also enhance the cross sections of 
$\Cha^+_i\Cha^-_{j}$ and $\Cha^0_i\Cha^0_{j} $ 
by about one order of magnitude and   the cross sections for the production of 
$\Cha^+_1\Cha^-_{1}$ and $\Cha^0_1\Cha^0_{2} $ could be up to $1$ pb. 
In addition, we also investigate the  processes that chargino and
neutralino are in the final state, e.g.  $pp\to \Cha^0_i\Cha^\pm_{j}$. 
The production mechanism proceeds via the conventional
Drell-Yan processes with W gauge boson,  charged Higgs boson and charged 
Goldstone. The dominant contribution is through $W$ gauge boson exchange.
The charged Higgs contribution is  through $c\bar{b}\to H^{\pm *} \to
\Cha^+_i\Cha^0_{j}$ and  the enhancement of large $\tan\beta$ is  from the 
bottom Yukawa coupling.  Unfortunately,  it turns out that this 
large $\tan\beta$ enhancement can not overcome the suppression from 
$V_{cb}$ Cabibbo-Kobayashi-Maskawa (CKM) matrix element.

\begin{figure}[t] 
\begin{picture}(320,260)
\put(-100,-250){\mbox{\psfig{file=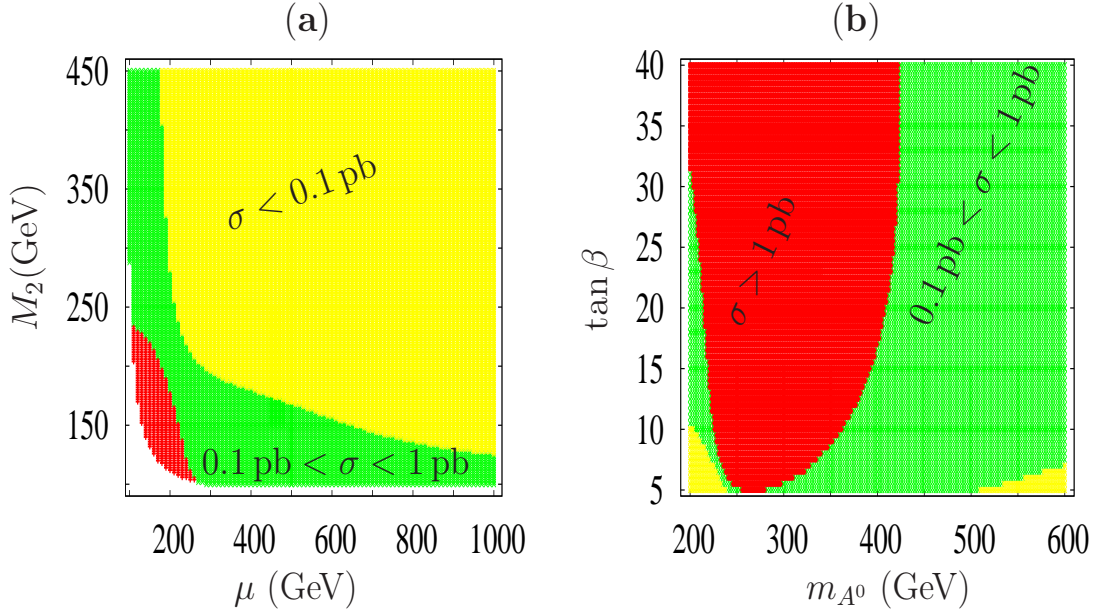,width=8.in,height=11in}}}
\end{picture}
\vspace{-0.1cm}
\caption{Scatter plots of $\sigma(pp\to b\bar{b}\to
\widetilde{\chi}^-_1 \widetilde{\chi}^+_1)$ in the 
 ($\mu, M_2$) plan (left) and ($m_{A^0}, \tan\beta$) plan (right). 
The SUSY parameters are chosen to be 
$M_{SUSY} = 490$ GeV, $ A_t = A_b = 1140$ GeV,
($m_{A^0} = 350\, {\rm GeV}, \tan\beta = 20 $) and ($M_2 = \mu = 150\, {\rm GeV}$) for left and right panels, respectively .
} 
\label{fig-c11}
\end{figure}
\begin{figure}[t] 
\begin{picture}(320,260)
\put(-100,-250){\mbox{\psfig{file=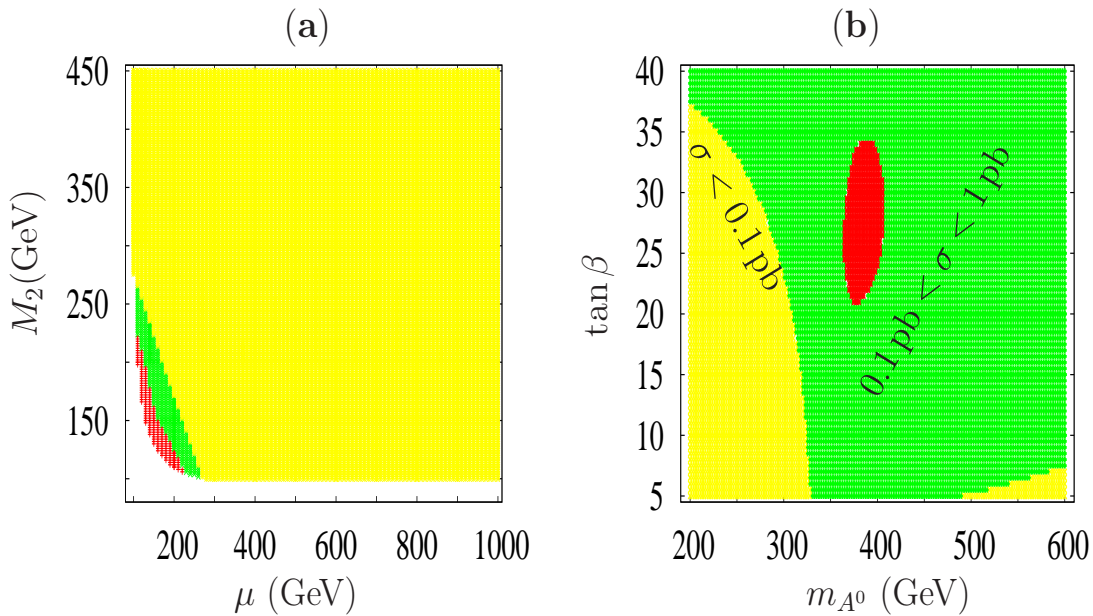,width=8.in,height=11in}}}
\end{picture}
\vspace{-0.1cm}
\caption{ Scatter plots of $\sigma(pp\to b\bar{b}\to\widetilde{\chi}^-_1 \widetilde{\chi}^+_2 + c.c)$ in the 
 ($\mu, M_2$) plan (left) and ($m_{A^0}, \tan\beta$) plan (right). The other
  parameters are fixed as in Fig.(\ref{fig-c11}) 
}
\label{fig-c12}
\end{figure}

\begin{figure}[t] 
\begin{picture}(320,260)
\put(-100,-250){\mbox{\psfig{file=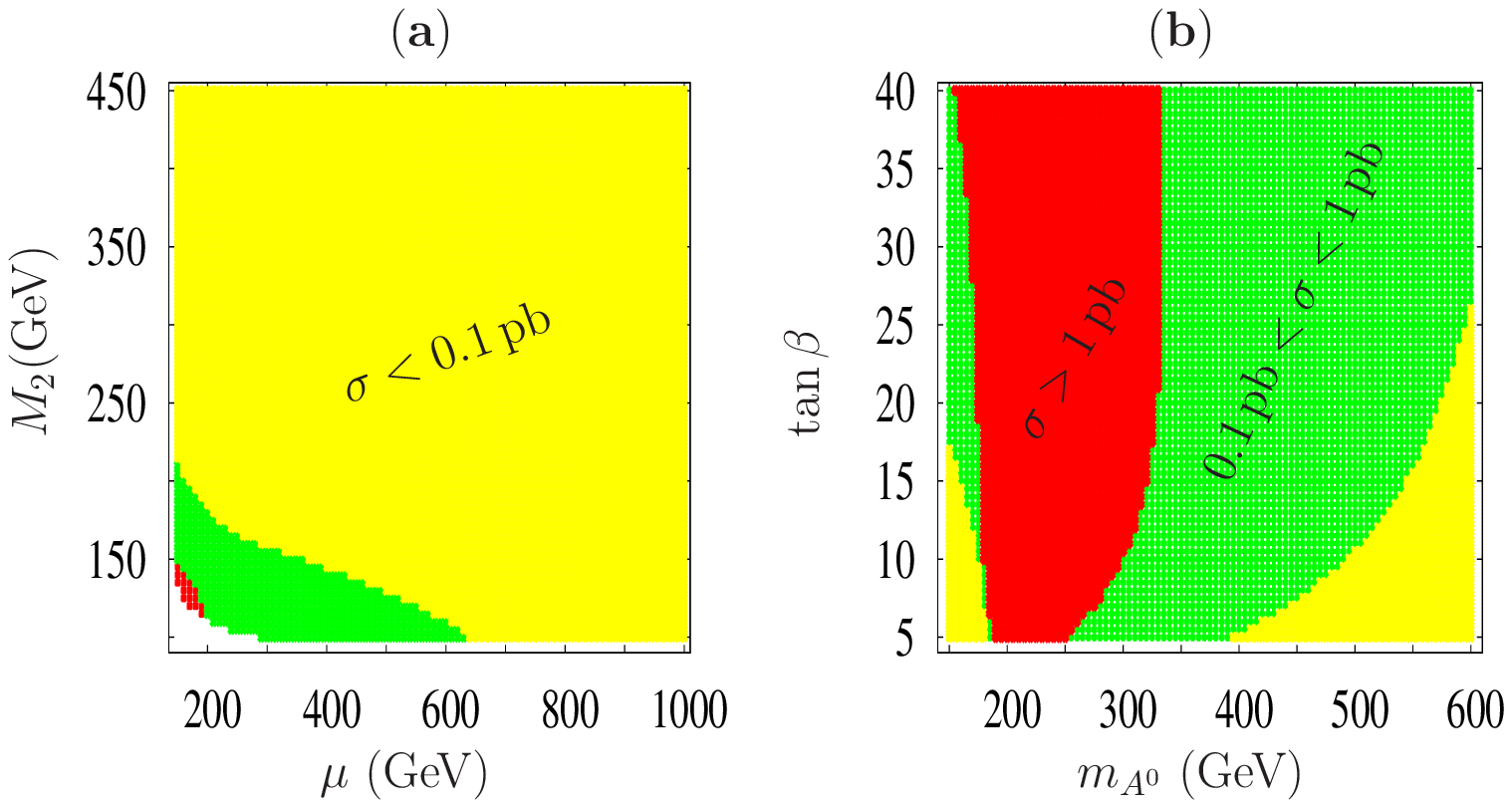,width=8.in,height=11in}}}
\end{picture}
\vspace{-0.1cm}
\caption{Scatter plots of $\sigma(pp\to b\bar{b}\to 
\widetilde{\chi}^0_1 \widetilde{\chi}^0_2)$ in the 
 ($\mu, M_2$) plan (left) and ($m_{A^0}, \tan\beta$) plan (right). 
The SUSY parameters are chosen to be ($m_{A^0} = 220\, 
{\rm GeV}, \tan\beta = 20 $) and ($M_2 = 120 $ GeV, 
$\mu = 150 $ GeV) for left and right panels, respectively.} 
\label{fig-n12}
\end{figure}


To quantify those effects from s-channel Higgs exchange contribution
 and to show their importance,
we provide some scatter plots in $(\mu,M_2)$ and $(\tan\beta,m_{A^0})$ plans.
 From the results in Figs.~\ref{fig-c11}(a) and \ref{fig-c12}(a),
we see that  with 
$\mu\la 250$ GeV and $M_2\la 250$ GeV,  since we are very close to the
resonant region, the cross section is slightly larger than
$1$pb. In the case of diagonal production of $\Cha^+_1\Cha^-_1$,  the region with large $\mu$ and moderate $M_2$ (gaugino like), or 
large $M_2$ and moderate $\mu$ (higgsino like) is interesting (see Fig.~\ref{fig-c11}(a)). This is because 
the process is dominated by the s-channel $Z^0$ exchange and 
the cross section can be in the range $0.1-1$pb. 
Due to the phase space suppression,  non diagonal production 
$\Cha^+_1\Cha^-_2$ will be small in this region. 
On the other hand we show in Fig.~\ref{fig-c11}(b) and Fig.~\ref{fig-c12}(b)
 the production cross section in the plan $(\tan\beta,m_{A^0})$.
Here we can see the resonant effect for $\Cha^+_1\Cha^-_1$ 
when $m_{A^0}\approx m_{H^0}\approx 280$ GeV. This effect is amplified for large 
$\tan\beta$. There is also a large area where the diagonal production cross
section  $\Cha^+_1\Cha^-_1$ is in the range $0.1-1$pb.
In the case of non diagonal production $\Cha^+_1\Cha^-_2$ and due to phase
space suppression the resonance effect is rather mild. That is the reason 
why one can see only small region for $\tan\beta\in [20,35]$ where the cross
section is larger than 1pb.

In the case of the associate production $\Cha^0_1\Cha^0_2$ 
we show the scatter plots in Fig.~\ref{fig-n12} in $(\mu,M_2)$
 and $(m_{A^0}, \tan\beta)$ plans. When $|\mu|\gg M_2$ the two lightest 
neutralinos are   both nearly pure gauginos, their s-channel contribution 
is then small, the squarks exchange diagrams play the most
   important role in this case.
Unlike $\Cha^+_1\Cha^-_2$ which suffers phase space
 suppression, $\Cha^0_1\Cha^0_2$ does not have such suppression. 
This is mainly due to the fact that  $m_{\chi_2^0}\approx m_{\chi_1^\pm}$ 
(see section II.A). 
Therefore, we can see in Fig.~\ref{fig-n12}(b) the same resonance effect
we have seen in the case of $\Cha^+_1\Cha^-_1$. 

Finally, in table~\ref{tab:separate} we give separate contributions 
to $b\bar{b}$ and gluon gluon fusion  
that originate from s-channel of Higgs $A^0$ and $H^0$ exchange only and also 
from the full  set of Feynman diagrams. 
It is clear from this table that s-channel Higgs exchange contribution 
is the dominant one. This can be viewed as a production of the Heavy Higgs 
bosons followed by the subsequent decays into a chargino or neutalino pairs
\cite{Bisset:2007mi,Baer:1992kd}.

\begin{table*}[h!]
\begin{centering}
\begin{ruledtabular}
\begin{tabular}{lc|cccc|cccc}
\multicolumn{2}{c|}{} & \multicolumn{4}{c|}{$\sqrt{s} = 14 $ TeV} &
\multicolumn{4}{c}{$\sqrt{s} = 7$ TeV} \tabularnewline
\multicolumn{2}{c|}{$\sigma$ {[}pb{]}} %
  & Higgs  Only   &   &   & Full  & Higgs Only &  &   & Full  \\
\hline
& $b\bar{b}\to \widetilde{\chi}^+_1\widetilde{\chi}^-_1$ 
& 3.761  &   &   &
3.881 & 0.727  &   &   & 0.755  \tabularnewline
& $b\bar{b}\to \widetilde{\chi}^+_1\widetilde{\chi}^-_2 + {\rm h.c}$ 
& 0.498  &   &   &
0.504 & 0.072  &   &   & 0.074 \tabularnewline
& $b\bar{b}\to \widetilde{\chi}^+_2\widetilde{\chi}^-_2$ 
& 0.054  &   &   &
0.066 & 0.006  &   &   & 0.007  \tabularnewline
& $ gg \to \widetilde{\chi}^+_1\widetilde{\chi}^-_1$ 
& 0.134  &   &   &
0.149 & 0.122  &   &   & 0.138  \tabularnewline
& $ gg \to \widetilde{\chi}^+_1\widetilde{\chi}^-_2 + {\rm h.c}$ 
& 0.086  &   &   &
0.098 & 0.081  &   &   & 0.019 \tabularnewline
& $ gg \to \widetilde{\chi}^+_2\widetilde{\chi}^-_2$ & 0.043  &   &   &
0.015 & 0.001  &   &   & 0.002  \tabularnewline
\hline
& $b\bar{b}\to \widetilde{\chi}^0_1\widetilde{\chi}^0_2 $ & 2.822  &   &   &
2.750 & 0.645  &   &   & 0.630 \tabularnewline
& $b\bar{b}\to \widetilde{\chi}^0_2\widetilde{\chi}^0_2$ & 1.384  &   &   &
1.333 & 0.285  &   &   & 0.276  \tabularnewline
& $ gg \to \widetilde{\chi}^0_1\widetilde{\chi}^0_2$ & 0.805  &   &   &
0.789 &  0.241 &   &   & 0.205  \tabularnewline
& $ gg \to \widetilde{\chi}^0_2\widetilde{\chi}^0_2 $ & 0.301  &   &   &
0.272 & 0.071  &   &   & 0.064 \tabularnewline

\end{tabular}
\end{ruledtabular}
\par\end{centering}
\caption{The
effect of the s-channel Higgs ($H^0, A^0$) on the production cross sections (in pb). The
SUSY parameters are chosen to be $A_t = A_b =  1140 $ GeV, $\mu = 150 $ GeV,
$M_2 = 120$ GeV, $M_{SUSY} = 490 $ GeV, $m_{\widetilde{g}} = 1$ TeV,
$\tan\beta = 40$ and the Higgs masses are taken at the resonance.}
\label{tab:separate}
\end{table*}

\section{Conclusion}
 We have studied the pair production of charginos and neutralinos in detail
where the study includes the  tree level s-channel Higgs bosons exchange 
and  the radiative corrections to the bottom Yukawa couplings. 
It has been shown that the s-channel Higgs bosons effect can enhance 
substantially the production cross section in the mixed region when $M_2$ and
$|\mu|$ are comparable and below 1 TeV. Such enhancement can go up to
one order of magnitude compared to the usual $q\bar{q}$ fusion contribution.
We have demonstrated that the enhancement has two origins: on one hand the 
large $\tan\beta$ enhancement and on the other hand resonance effect from
s-channel Higgs bosons. 
Such enhancements exceed the PDF uncertainties on the evaluation of the cross
section and are  in some case larger than the NLO correction. Therefore, these
contributions have to be taken into account in any reliable future analysis.
We have found that in the low $\tan\beta$ regime, the gluon fusion 
 contribution could be comparable to $q\bar{q}$ and $b\bar{b}$ one.
 Those processes can be used to extract some
 information on the chargino neutralino Higgs couplings right at the Higgs
 boson resonances and the involved SUSY
 parameters.

\section{ACKNOWLEDGEMENTS}
We thank FEDERICO von der PAHLEN for helpful discussions.  
A.A is supported by the NSC 
under contract  \# 100-2811-M-006-008. The work of R.B was supported by
CSIC. M.C. would like to thank C-H. Chen and H-N. Li for invitation and
hospitality at NCKU and Academia Sinica and acknowledge NCTS support. 
CHC was supported by NSC Grant No.NSC-97-2112-M-006-001-MY3.


\appendix

\section{SUSY couplings}
We describe in this appendix all the couplings of 
these SUSY particles i.e. 
couplings of the neutralinos and charginos to gauge and Higgs bosons and their
couplings to fermion--sfermion pairs as well as the couplings of MSSM Higgs
and gauge bosons to fermions, which will be needed later when evaluating the
cross sections of $2\to 2$ processes. We will use the notation of~
\cite{Haber:1984rc, Gunion:1984yn}
\subsection{Chargino and Neutralino Interactions}
We start this section by discussing the chargino and neutralino interactions
with gauge bosons $(\gamma, Z$ and $W^\pm )$, Higgs bosons as well as fermion-sfermions pairs.\\
The resulting charged and neutral 
weak boson terms in the Lagrangian density, 
expressed in the four
component notation and in the weak basis reads
\begin{eqnarray}
{\cal{L}} &=& - eA_\mu \overline{\Cha^+_k}\gamma^\mu \Cha^+_k + 
\frac{g}{c_W} Z_\mu \sum_{\al,m,k} \overline{\Cha^+_m}\gamma^\mu {\cal{O}}^\al_{mk} \, P_\al \, \Cha^+_{k} \\ &+& \frac{g}{2\,\,c_W} Z_\mu \sum_{\al, l,n} \overline{\Cha^0_l}\gamma^\mu {\cal{N}}^\al_{ln}\,P_\al \,\Cha^0_{n}
+ \big[g\,W^-_{\mu}\sum_{\al,l,k} \overline{\Cha^0_l}\gamma^\mu {\cal{C}}^\al_{lk}\,P_{\al}\, \Cha^+_{k} + {\rm H.m}\big]\non
\end{eqnarray}
where $g = e/s_W$, $k, m =$ 1,2 for the chargino and $l, n = $1,...4 for the neutralino, $\al = L, R$ with $P_{L, R} = (1 \mp \gamma_5)/2$.
The couplings ${\cal{O}}^{\al}_{mk}$, ${\cal{N}}^{\al}_{ln}$ and ${\cal{C}}^{\al}_{lk}$ are given by
\begin{eqnarray}
{\cal{O}}^{L}_{mk} &=& - V_{m1} V^*_{k1} - \frac{1}{2}V_{m2} V^*_{k2} + \delta_{mk} s^2_W,\label{2}\\
{\cal{O}}^{R}_{mk} &=& - U^*_{m1} U_{k1} - \frac{1}{2}U^*_{m2} U_{k2} + \delta_{mk} s^2_W,\\
{\cal{N}}^{L}_{ln} &=& -\frac{1}{2} Z_{l3} Z^*_{n3} + \frac{1}{2} Z_{l4} Z^*_{n4},\\
{\cal{N}}^{R}_{ln} &=& - \big({\cal{N}}^{L}_{ln}\big)^*,\\
{\cal{C}}^{L}_{lk} &=& - \frac{1}{\sqrt{2}}Z_{l4} V^*_{k2} + Z_{l2} V^*_{k1},\\
{\cal{C}}^{R}_{lk} &=& \frac{1}{\sqrt{2}}Z^*_{l3} U_{k2} + Z^*_{l2} U_{k1}\label{7}.
\end{eqnarray}
$Z$, $U$, $V$ are the neutralino and chargino mixing matrices, respectively. 
The unitarity properties of the $U$ and $V$ matrices have been used in deriving
Eqs.~(\ref{2})-(\ref{7}).
\par
The couplings of the Higgs bosons to the electroweak neutralinos 
and charginos originate from the gauge strength Yukawa couplings 
of gauginos to the scalar and fermionic components 
of a given chiral supermultiplet. 
In four-conponent the Lagrangian reads as:
\begin{eqnarray}
{\cal{L}}  &=& - \frac{g}{2} \sum_{i=1,2} H^0_i \overline{\Cha^0_l} 
S_{lni}\Cha^0_n - \frac{g}{2} \sum_{i=3,4} H^0_i \overline{\Cha^0_l} 
S_{lni}\gamma_5 \Cha^0_n\\\non
&-& g \sum_{i=1,2} H^0_i \overline{\Cha^+_k}
\left( C_{kmi}\, P_R + C^*_{mki}\,P_L \right)\Cha^+_m + 
ig\sum_{i=3,4} H^0_i \overline{\Cha^+_k}\left( C_{kmi}\, 
P_R + C^*_{mki}\,P_L \right)\Cha^+_m\\\non
&-& g \sum_{i=1,2} \big [ H^+_i \overline{\Cha^+_k}
\left( F^R_{kli}\, P_R + F^L_{kli}\,P_L \right)\Cha^0_l + {\rm H.c}\big]
\end{eqnarray}
where the couplings are given by:
\begin{eqnarray}
S_{lni} &=& \frac{e_i}{2} \bigg [ Z_{l3} Z_{n2} + Z_{n3} Z_{l2} - \tan\theta_W 
(Z_{l3} Z_{n1} + Z_{n3} Z_{l1})\bigg]\\\non
&+& \frac{d_i}{2} \bigg [ Z_{l4} Z_{n2} + Z_{n4} Z_{l2} - \tan\theta_W 
(Z_{l4} Z_{n1} + Z_{n4} Z_{l1})\bigg],\\
C_{kmi} &=& \frac{1}{\sqrt{2}} \left( e_i V_{k1}U_{m2} - d_i V_{k2} U_{m1}\right)\\
C^*_{mki} &=& C_{kmi} \ \ {\rm for } \ \ i =1,2 \ \ {\rm and} \ \ C^*_{mki} = - C_{kmi} \ \ {\rm for } \ \ i =3,4\\
F^R_{kli} &=& d_{i+2} \bigg[ V_{k1} Z_{l4} +\frac{1}{\sqrt{2}} (Z_{l2} + Z_{l1}
\tan\theta_W)V_{k2}\bigg]\\
F^L_{kli} &=& -e_{i+2} \bigg[ U_{k1} Z_{l3} -\frac{1}{\sqrt{2}} (Z_{l2} + Z_{l1}
\tan\theta_W)U_{k2}\bigg]
\end{eqnarray}
Again, here we have used $k, m =$ 1,2 for the chargino and $l, n = $1,...4 for the neutralino. $H^0_i=(h^0, H^0, A^0, G^0)$ ($i$=1...4), and $H^+_i=(H^+, G^+)$ ($i$=1,2) $d_i$ and $e_i$ take the values
\begin{eqnarray}
d_i = \bigg(-\cos\alpha, -\sin\alpha, \cos\beta, \sin\beta\bigg), \quad e_i = \bigg(-\sin\alpha, \cos\alpha, -\sin\beta, \cos\beta\bigg)
\end{eqnarray}

\par
The squark-quark-chargino Lagrangian is given by,
\begin{eqnarray} 
{\cal{L}} &=& g\bigg[\overline{u}A^L_{sk} \, P_R \tilde{d}_s \Cha^+_{k} + 
\tilde{d}^{\dagger}_s\, \overline{\Cha^+_{k}} \,B^L_{sk} \, P_R u
+ \overline{d}E^R_{sk} \, P_R \tilde{u}_s (\Cha^+_{k})^C \\\non
&+& \tilde{u}^{\dagger}_s\, (\Cha^+_{k})^C \,F^L_{sk} \, P_L \, d \bigg]
+ {\rm H.c}
\end{eqnarray} 
with the following couplings
\begin{eqnarray}
A^L_{sk} &=& -V_{ud} \bigg[ U^*_{k1} R^{\widetilde{d}}_{s1} - \frac{m_d}{\sqrt{2} M_W c_\beta} U^*_{k2} R^{\widetilde{d}}_{s2}\bigg]\\
B^L_{sk} &=& \frac{m_u}{\sqrt{2} M_W s_\beta} V_{k2}\,R^{\widetilde{d}}_{s1} V_{ud} \\
E^R_{sk} &=& -V_{ud} \bigg[ V^*_{k2} R^{\widetilde{u}}_{s2} - \frac{m_u}{\sqrt{2} M_W s_\beta} V^*_{k1} R^{\widetilde{u}}_{s1}\bigg]\\
F^L_{sk} &=& \frac{m_d}{\sqrt{2} M_W c_\beta} U^*_{k2}\,R^{\widetilde{u}}_{s1} V_{ud} \\\non
\end{eqnarray}
$R^{\widetilde{d},\widetilde{u}}_{ss^\prime}$ with ($s, s^\prime$ = 1,2) are the elements of the rotation matrices diagonalizing the up- and down- type squark mass matrices, and $V_{ud}$ are the elements of the CKM matrix. 
The squark-quark-neutralino interaction can be 
written down in a similar way,
\begin{eqnarray} 
{\cal{L}}  = g\,\overline{\Cha^0_l} \bigg[ (G^{uL}_{isl} P_L + G^{uR}_{isl} P_R)\widetilde{u}^{\dagger}_s u_i + 
(G^{dL}_{isl} P_L + G^{dR}_{isl} P_R)\widetilde{d}^{\dagger}_s d_i \bigg]
+ {\rm H.c}
\end{eqnarray}
where the couplings are defined as
\begin{eqnarray}
G^{uL}_{sl} &=& \sqrt{2} e_u \tan\theta_W R^{\widetilde{u}}_{s2} Z^*_{n1} - \frac{m_u}{\sqrt{2}\,M_W s_\beta} R^{\widetilde{u}}_{s1} Z^*_{n4}\\
G^{uR}_{sl} &=& -\frac{m_u}{\sqrt{2}M_W s_\beta} R^{\widetilde{u}}_{s2} Z_{n4} - 
\frac{e_u (s_W Z_{n1} + 3c_W Z_{n2})}{2\sqrt{2}c_W} R^{\widetilde{u}}_{s1}\\
G^{dL}_{sl} &=& \sqrt{2} e_d \tan\theta_W R^{\widetilde{d}}_{s2} Z^*_{n1}
- \frac{m_d}{\sqrt{2}M_W c_\beta} R^{\widetilde{d}}_{s1} Z^*_{n3}\\
G^{dR}_{sl} &=& -\frac{m_d}{\sqrt{2}M_W c_\beta} R^{\widetilde{d}}_{s2} Z_{n3} + 
\frac{e_d (s_W Z_{n1} - 3c_W Z_{n2})}{2\sqrt{2}c_W} R^{\widetilde{d}}_{s1}\\\non
\end{eqnarray}
with $e_{u} = 2/3 $ and $e_d= -1/3$.\\

\section{Production rates}
\label{cross-sections}

The production of chargino/neutralino pair, 
as initiated by $b\bar{b}$ annihilation, involves photon, $Z$ and Higgs bosons 
in the $s$-channel as well as squark/slepton exchanges in the $t$/$u$-channels. 
We present the differential cross section for each subprocess
separately in the mass eigen-basis. The summation and average of
spin/color for final and initial states are taken into account. In the
formulas presented below, summation over repeated indices $k$ and
$k^\prime$ for the Higgs bosons and $s$ and $l$ for the squark and
sleptons in the intermediate states are understood. 
Now let us define our notation for the convenience of the following formulas.  
The momenta of the incoming quark $b$ and anti-quark $\bar b$, 
outgoing $\widetilde{\chi}_i$ and outgoing $\widetilde{\chi}_j$
are denoted by $p_1$, $p_2$, $k_1$ and $k_2$, respectively.
We neglect the quark masses of the incoming partons.  
The Mandelstam variables are defined as follows:
\begin{eqnarray}
\hat s &=&  (p_1+p_2)^2 = (k_1+k_2)^2  \nonumber \\
\hat t &=& (p_1-k_1)^2 = (p_2-k_2)^2 
= \frac{ m_{\wt{\chi}_i}^2 + m_{\wt{\chi}_j}^2}{2} - 
\frac{\hat s}{2} \left( 1 - \beta \cos\theta^* \right ) \nonumber \\
\hat u &=& (p_1-k_2)^2 = (p_2-k_1)^2 
= \frac{ m_{\wt{\chi}_i}^2 + m_{\wt{\chi}_j}^2}{2} - 
\frac{\hat s}{2} \left( 1 + \beta \cos\theta^* \right ) 
\label{mandelstan}
\end{eqnarray}
where $\beta = \lambda^{1/2}(1, m_{\wt{\chi}_i}^2/\hat s, m_{\wt{\chi}_j}^2/\hat s)$ 
and $\theta^*$ is
the scattering angle in the center-of-mass frame of the partons.
\section*{Chargino-pairs production}
\begin{eqnarray}
\frac{d\hat\sigma_{LO}}{d\hatt}(\Cha^+_i\Cha^-_j)\non &=& 
\frac{4\pi^2\,\alpha^2}{3s^4_W}
\bigg[\Bigg(
\frac{8\,e^2_q}{\hats^2}\Big[\hats^2 + 2\big( m^4_{\Cha^\pm_i} - 2m^2_{\Cha^\pm_i}\hatt + \hats\hatt + \hatt^2\big)\Big] 
-\frac{4 s^2_W\,e_q}{c_W } \frac{D_Z}{\hats} \Big[ g_{RZ} (\hatt^2 
\\\non &+& 
m^4_{\Cha^\pm_i} - 2 m^2_{\Cha^\pm_i}\hatt - m^2_{\Cha^\pm_i}\hats) ({\cal{O}}^L_{ij} + {\cal{O}}^R_{ij}) + {\cal{O}}^R_{ij} (\hats^2 + 2\hats\hatt) + 
(L\leftrightarrow R)\Big]
\\\non&+&
4\sqrt{2}\,e_q\,s^2_W \frac{D_{\phi}}{\hats}C_{iik}\,g_{bb\phi}\,m_b\,m_{\Cha^\pm_{i}}\Big(\hats + 2\hatt - 2m^2_{\Cha^\pm_{i}}\Big)-\frac{e_q\,s^2_W}{\hats} \Big((E^R_{si})^2 \\\non &+& (F^L_{si})^2\Big)
\Big[m^4_{\Cha^\pm_{i}} + (\hats + \hatt)^2 - m^2_{\Cha^{\pm}_{j}}(\hats + 2\hatt)\Big]U_{\widetilde{t}_s}\Bigg)\delta_{ij}
+  
\frac{D^2_{Z}}{c^2_W}
\Big[ 
g^2_{RZ} 
\Big ( 2 m_{\Cha^\pm_i}\\\non &\times& m_{\Cha^\pm_j} {\cal{O}}^{L}_{ij}{\cal{O}}^{R}_{ij} \hats 
+ ({\cal{O}}^{L}_{ij})^2 \hats^2  
+ (({\cal{O}}^{L}_{ij})^2  + ({\cal{O}}^{R}_{ij})^2 )\hatt^2 
+2\,({\cal{O}}^{L}_{ij})^2 \hats \hatt +
m^2_{\Cha^\pm_i} \\\non &\times&\Big[ m_{\Cha^\pm_j} (({\cal{O}}^{L}_{ij})^2  + ({\cal{O}}^{R}_{ij})^2 ) - ({\cal{O}}^{R}_{ij})^2 \hatt - ({\cal{O}}^{L}_{ij})^2 (\hats + \hatt)\Big]
-m^2_{\Cha^\pm_j} (({\cal{O}}^{R}_{ij})^2 \hatt \\\non &+& ({\cal{O}}^{L}_{ij})^2 (\hatt+\hats))\Big) + (L\leftrightarrow R)\Big]
+ g^2_{bb\phi} D^2_{\phi} \hats \Big[
(C^2_{ij\phi} + C^2_{ji\phi})(\hats - m^2_{\Cha^\pm_i} \\\non &-& m^2_{\Cha^\pm_j}) - 4\,m_{\Cha^\pm_i}m_{\Cha^\pm_j} C_{ij\phi}C_{ji\phi}\Big]
+g^2_{bbA} D^2_{A} \hats \Big[
(C^2_{ijA} + C^2_{jiA})(\hats - m^2_{\Cha^\pm_i} \\\non &-& m^2_{\Cha^\pm_j}) + 4\,m_{\Cha^\pm_i}m_{\Cha^\pm_j} C_{ijA}C_{jiA}\Big]
 + ((E^R_{si})^2 + (F^L_{si})^2)((E^R_{sj})^2 + (F^L_{sj})^2)\\\non&\times& 
(\hatt - m^2_{\Cha^\pm_i})(\hatt - m^2_{\Cha^\pm_j}) T^2_{\widetilde{t}_s}
+\frac{\sqrt{2}}{c_w}g_{bb\phi}D_{\phi}D_{Z}\,m_{b} (\hats + 2\hatt - m^2_{\Cha^\pm_i} - m^2_{\Cha^\pm_i})\\\non &\times&
\Big[(m_{\Cha^\pm_i} C_{ji\phi} + m_{\Cha^\pm_j} C_{ij\phi}){\cal{O}}^L_{ij} +
(m_{\Cha^\pm_i} C_{ij\phi} + m_{\Cha^\pm_j} C_{ji\phi}){\cal{O}}^R_{ij}
\Big]
\\\non &-&
\frac{\hats\,m_b}{\sqrt{2}c^2_WM^2_Z} g_{bbA}D_{A}D_{Z}
\Big[
(m^2_{\Cha^\pm_i} - m^2_{\Cha^\pm_j})\Big(
C_{ijA}(m_{\Cha^\pm_j}{\cal{O}}^L_{ij} + m_{\Cha^\pm_i}{\cal{O}}^R_{ij})
\\\non &-& 
C_{jiA}(m_{\Cha^\pm_i}{\cal{O}}^L_{ij} + m_{\Cha^\pm_j}{\cal{O}}^R_{ij})
\Big)+
\hats \Big(
C_{jiA}(m_{\Cha^\pm_i}{\cal{O}}^L_{ij} - m_{\Cha^\pm_j}{\cal{O}}^R_{ij})
\\\non &+&  
C_{ijA}(m_{\Cha^\pm_j}{\cal{O}}^L_{ij} - m_{\Cha^\pm_i}{\cal{O}}^R_{ij})
\Big)
\Big]-
\frac{2}{c_W} D_Z T_{\widetilde{t}_s}
\Big[ m_{\Cha^\pm_i}m_{\Cha^\pm_j} \Big( F^L_{si} F^L_{sj}g_{RZ}
\\\non &\times& (m_{\Cha^\pm_i}m_{\Cha^\pm_j}{\cal{O}}^R_{ij} 
+ 
\hats {\cal{O}}^L_{ij} ) + E^R_{si} E^R_{sj} g_{LZ} 
(m_{\Cha^\pm_i}m_{\Cha^\pm_j}{\cal{O}}^L_{ij} + \hats {\cal{O}}^R_{ij} )
-(\hatt \\\non &-&  m^2_{\Cha^\pm_i} -  m^2_{\Cha^\pm_j})
\Big(
E^R_{si} E^R_{sj} g_{LZ}{\cal{O}}^L_{ij} + 
F^L_{si} F^L_{sj} g_{RZ}{\cal{O}}^R_{ij}
\Big)\hatt
+
2g_{bbh}g_{bbH}\hats D_{h}D_{H}\\\non &\times&\Big[
2m_{\Cha^\pm_i}m_{\Cha^\pm_j} (C_{ijh}C_{jiH} +  C_{ijH}C_{jih}) 
+(m^2_{\Cha^\pm_i} +m^2_{\Cha^\pm_j} - \hats)
(C_{ijh}C_{ijH} \\\non &+&  C_{jih}C_{jiH})
-\sqrt{2}g_{bb\phi}D_{\phi}T_{\widetilde{t}_s} 
(C_{ij\phi} E^R_{sj}F^L_{si} + C_{jih}E^R_{si}F^L_{sj})(m_{\Cha^\pm_i}m_{\Cha^\pm_j} + \hatt)\hatt
\\ &+&
2 (E^R_{si} E^R_{s^\prime i} + F^L_{si}F^L_{s^{\prime}i})(E^R_{sj} E^R_{s^\prime j} + F^L_{sj}F^L_{s^{\prime}j})(\hatt - m^2_{\Cha^\pm_i})(\hatt - m^2_{\Cha^\pm_j}) T_{\widetilde{t}_s} T_{\widetilde{t}_{s^\prime}}  
\bigg]
\end{eqnarray}

\section*{Neutralino-pairs production}
\begin{eqnarray}
\frac{d\hat\sigma_{LO}}{d\hatt}(b\bar{b}\to \Cha^0_n\Cha^0_l)\non &=& \bigg(\frac{1}{1+\delta_{nl}}\bigg)
\frac{4\pi^2\alpha^2}{3s^4_W} \Bigg[\frac{(g^2_{LZ} + g^2_{RZ})}{c^2_W}(N^R_{nl})^2 D^2_Z \Big( 
2m^2_{\Cha^0_n}m^2_{\Cha^0_l} + \hats^2 + 2\hats\hatt + 2\hatt^2 \\\non &-& 
(m^2_{\Cha^0_n}+ m^2_{\Cha^0_l})(\hats + 2\hatt) 
- 2m_{\Cha^0_n}m_{\Cha^0_l}\hats 
\Big) + g^2_{bbh}D^2_{h}S^2_{nlh}\Big(
\hats^2 - 2m_{\Cha^0_n}m_{\Cha^0_l}\hats \\\non &-& (m^2_{\Cha^0_n} + m^2_{\Cha^0_l})\hats + g^2_{bbA}D^2_{A}S^2_{nlA}\Big(
\hats^2 + 2m_{\Cha^0_n}m_{\Cha^0_l}\hats - (m^2_{\Cha^0_n} + m^2_{\Cha^0_l})\hats
\Big)\\\non &+& \Big( (G^{dL}_{sl})^2 + (G^{dR}_{sl})^2 \Big)\Big( (G^{dL}_{sn})^2 + (G^{dR}_{sn})^2 \Big)(m^2_{\Cha^0_n} - \hatt)(m^2_{\Cha^0_l} - \hatt)T^2_{\widetilde{b}_s}\\\non &+& \Big( (G^{dL}_{sl})^2 + (G^{dR}_{sl})^2 \Big)\Big( (G^{dL}_{sn})^2 + (G^{dR}_{sn})^2 \Big)(m^2_{\Cha^0_n} - \hat{u})(m^2_{\Cha^0_l} - \hat{u})U^2_{\widetilde{b}_s}
\\\non &+& \frac{2\,N^R_{nl}}{c_W}\Big( G^{dL}_{sl} G^{dR}_{sn} g_{RZ} - G^{dL}_{sn} G^{dR}_{sl} g_{LZ} \Big)D_Z T_{\widetilde{b}_s} \Big( (m^2_{\Cha^0_n} - \hatt)\hatt \\\non &+& m^2_{\Cha^0_l}(\hatt - m^2_{\Cha^0_n}) + m_{\Cha^0_n}m_{\Cha^0_l}\hats\Big)
+ \frac{2\,N^R_{nl}}{c_W}\Big( G^{dL}_{sl} G^{dR}_{sn} g_{RZ} - G^{dL}_{sn} G^{dR}_{sl} g_{LZ} \Big)  \\\non &\times& D_Z U_{\widetilde{b}_s}
\Big( (m^2_{\Cha^0_l} - \hat{u})(m^2_{\Cha^0_n}-\hat{u}) - m_{\Cha^0_n}m_{\Cha^0_l}\hats\Big)
- \frac{m_b (m_{\Cha^0_n} + m_{\Cha^0_l})}{M^2_W} \\\non &\times& D_Z D_A g_{bbA} S_{nlA} 
N^{R}_{nl}\Big( (m_{\Cha^0_l} - m_{\Cha^0_n})^2 - \hats\Big)\Big(2(g_{LZ}-g_{RZ})M_WM_Z - \hats \Big)
\\\non &-& 2 g_{bbh}g_{bbH} D_{h}D_{H}\Big(\hats^2 - (m^2_{\Cha^0_n} - m^2_{\Cha^0_l} ) \hats - 2 m_{\Cha^0_n} m_{\Cha^0_l}\hats\Big)+
g_{bbh} D_h S_{nlh} \hats \\\non &\times& \Big( G^{dR}_{sl} G^{dR}_{sl} + G^{dL}_{sn} G^{dL}_{sn}\Big)\Big( (\hatt + m_{\Cha^0_n}m_{\Cha^0_l})T_{\widetilde{b}_s} + (\hat{u} + m_{\Cha^0_n}m_{\Cha^0_l})U_{\widetilde{b}_s}\Big)\\\non &-&
g_{bbA} S_{nlA} D_{A} \hats \Big( G^{dR}_{sl} G^{dR}_{sl} + G^{dL}_{sn} G^{dL}_{sn}\Big) \Big((\hatt - m_{\Cha^0_n}m_{\Cha^0_l})T_{\widetilde{b}_s} \\\non &+& (\hat{u} - m_{\Cha^0_n}m_{\Cha^0_l})U_{\widetilde{b}_s} \Big) + 
2\,(G^{dL}_{sl}G^{dL}_{s^{\prime}l} + G^{dR}_{sl}G^{dR}_{s^{\prime}l})(G^{dL}_{sn}G^{dL}_{s^{\prime}n} + G^{dR}_{sn}G^{dR}_{s^{\prime}n})\\\non&\times&
\Big[(m^2_{\Cha^0_n} - \hatt)(m^2_{\Cha^0_l} - \hatt)T_{\widetilde{b}_s}T_{\widetilde{b}_{s^\prime}} + (m^2_{\Cha^0_n} - \hat{u})(m^2_{\Cha^0_l} - \hat{u})U_{\widetilde{b}_s}U_{\widetilde{b}_{s^\prime}}\Big]\\ &-&2\Big( m^2_{\Cha^0_n}m^2_{\Cha^0_l} {\cal{P}}_{nlss^{\prime}} + \hats m_{\Cha^0_n}m_{\Cha^0_l} {\cal{Q}}_{nlss^{\prime}} - \hatt\hat{u} {\cal{R}}_{nlss^{\prime}}\Big)T_{\widetilde{b}_s}U_{\widetilde{b}_{s^\prime}}
\Bigg]
\end{eqnarray}

with
\begin{eqnarray}
D_{Z} &=& \frac{1}{\hats - m^2_{Z} + i\,m_{Z}\Gamma_{Z}},\ \  D_{\phi} =  \frac{1}{\hats - m^2_{\phi} + i\,m_{\phi}\Gamma_{\phi}}, \quad {\rm with}\quad {\rm \phi = h^0,\, H^0,\, A^0}\\
T_{\widetilde{b}_s} &=& \frac{1}{\hatt - m^2_{\widetilde{b}_s}}, \quad U_{\widetilde{b}_s} = \frac{1}{\hat{u} - m^2_{\widetilde{b}_s}}.\\
{\cal{P}}_{nlss^{\prime}} &=& G^{dL}_{sn}G^{dL}_{sl}G^{dR}_{s^{\prime}l}G^{dR}_{s^{\prime}n} + G^{dL}_{s^{\prime}l}G^{dL}_{s^{\prime}n}G^{dR}_{sl}G^{dR}_{sn},\\
{\cal{Q}}_{nlss^{\prime}} &=& G^{dL}_{sn}G^{dL}_{s^{\prime}l}G^{dR}_{sl}G^{dR}_{s^{\prime}n} + G^{dL}_{sl}G^{dL}_{s^{\prime}l}G^{dR}_{sn}G^{dR}_{s^{\prime}n},\\
{\cal{R}}_{nlss^{\prime}} &=& G^{dL}_{s^{\prime}l}G^{dL}_{s^{\prime}n}G^{dR}_{sl}G^{dR}_{sn} + G^{dL}_{sl}G^{dL}_{sn}G^{dR}_{s^{\prime}l}G^{dR}_{s^{\prime}n}.\\\non
\end{eqnarray}

The factor $1/(1+\delta_{nl})$ is due to the two identical particles 
in the final states.


\end{document}